\documentclass[10pt,a4paper,twoside,twocolumn,english,prb]{revtex4}
\pdfoutput=1

\usepackage[T1]{fontenc}
\usepackage[latin9]{inputenc}
\usepackage{float}
\usepackage{amstext}
\usepackage{amssymb}
\usepackage{graphicx}
\usepackage{esint}
\usepackage{subscript}

\makeatletter
\usepackage{titlesec}

\makeatother

\usepackage{babel}

\newcommand{\citeabajo}[1]{~[\onlinecite{#1}]}

\newcommand{\nophase}{\orbital,\dwave,\stripes,\BEC}
\newcommand{\phase}{Franz,Emery,\Lascialfari,\Anderson,\Ong}

\newcommand{\phaseSinEmery}{Franz,\Lascialfari,\Anderson,\Ong}

\newcommand{\orbital}{VarmaNat}
\newcommand{\dwave}{Zhang}
\newcommand{\stripes}{\challengesstripes,\stripesSinChallenges}
\newcommand{\stripesSinChallenges}{KivelsonStr}
\newcommand{\BEC}{GIL}

\newcommand{\reconstruccion}{PRX,Cyr}

\newcommand{\Lascialfari}{Lascialfari2002}
\newcommand{\Anderson}{AndersonNat}
\newcommand{\Ong}{\OngNer,\OngMag}
\newcommand{\OngNer}{OngNeruno}
\newcommand{\OngMag}{OngMagPRB}

\newcommand{\challengesstripes}{PRX,Cyr}

\newcommand{\challengesMag}{Dancausa,CommentOng}

\begin{document}

\title{Critical temperatures for superconducting phase-coherence and condensation in La$_{2-x}$Sr$_{x}$CuO$_{4}$}

\author{Noelia Cot\'on, Manuel V. Ramallo, F\'elix Vidal \\ \it LBTS, Departamento
de F\'{\i}sica da Materia Condensada, Universidade de Santiago de Compostela,
ES-15782 Spain}

\maketitle

%%%%%%%%%%%%%%%%%%%%%%%%%%%%%%%%%%%%%%%%%%%%%%%%%%%%%%
%%%%%%%%%%%%%%%%%%%%%%%%%%%%%%%%%%%%%%%%%%%%%%%%%%%%%%
%%%%%%%%%%%%%%%%%%%%%%%%%%%%%%%%%%%%%%%%%%%%%%%%%%%%%%
%%%%%%%%%%%%%%%%%%%%%%%%%%%%%%%%%%%%%%%%%%%%%%%%%%%%%%

\noindent \textbf{A pivotal ongoing debate about cuprate superconductors (HTS) is the location of the transition temperatures for the superconducting wave function phase coherence and condensation, $T_{{\rm phase}}$ and $T_{{\rm cond}}$.\cite{Pines,NatPhysCoral,Franz,Tranquada} This shall elucidate which of two very different interactions dictate the macroscopic superconducting phase diagram of HTS: either those between normal-state carriers\cite{PRX,Cyr,KivelsonStr,VarmaNat,ShekhterRamshawetal2013,Zhang,GIL}
or those between pre-formed vortices and antivortices\cite{\phase}. Here, we present unambiguous experimental determinations of $T_{{\rm phase}}$ and $T_{{\rm cond}}$ in the prototypical HTS }La$_{2-x}$Sr$_{x}$CuO$_{4}$\textbf{ as a function of the doping level }\textbf{\textit{x}}\textbf{. $T_{{\rm phase}}$ is measured as a sharp change in the exponent $\alpha$ of the voltage-current characteristics $V\propto I^{\alpha}$. $T_{{\rm cond}}$ is determined from the critical rounding of the ohmic resistivity above $T_{{\rm phase}}$. Our measurements indicate that the transition to macroscopic superconductivity is accompanied by phase coherence due to vortex-antivortex binding and also that, for all }\textbf{\textit{x}}\textbf{, $T_{{\rm cond}}$ lies only a few Kelvin above $T_{{\rm phase}}$, limiting then the shift of the transition due to vortex-antivortex correlations.}

\begin{figure}
\includegraphics[width=1.05\columnwidth]{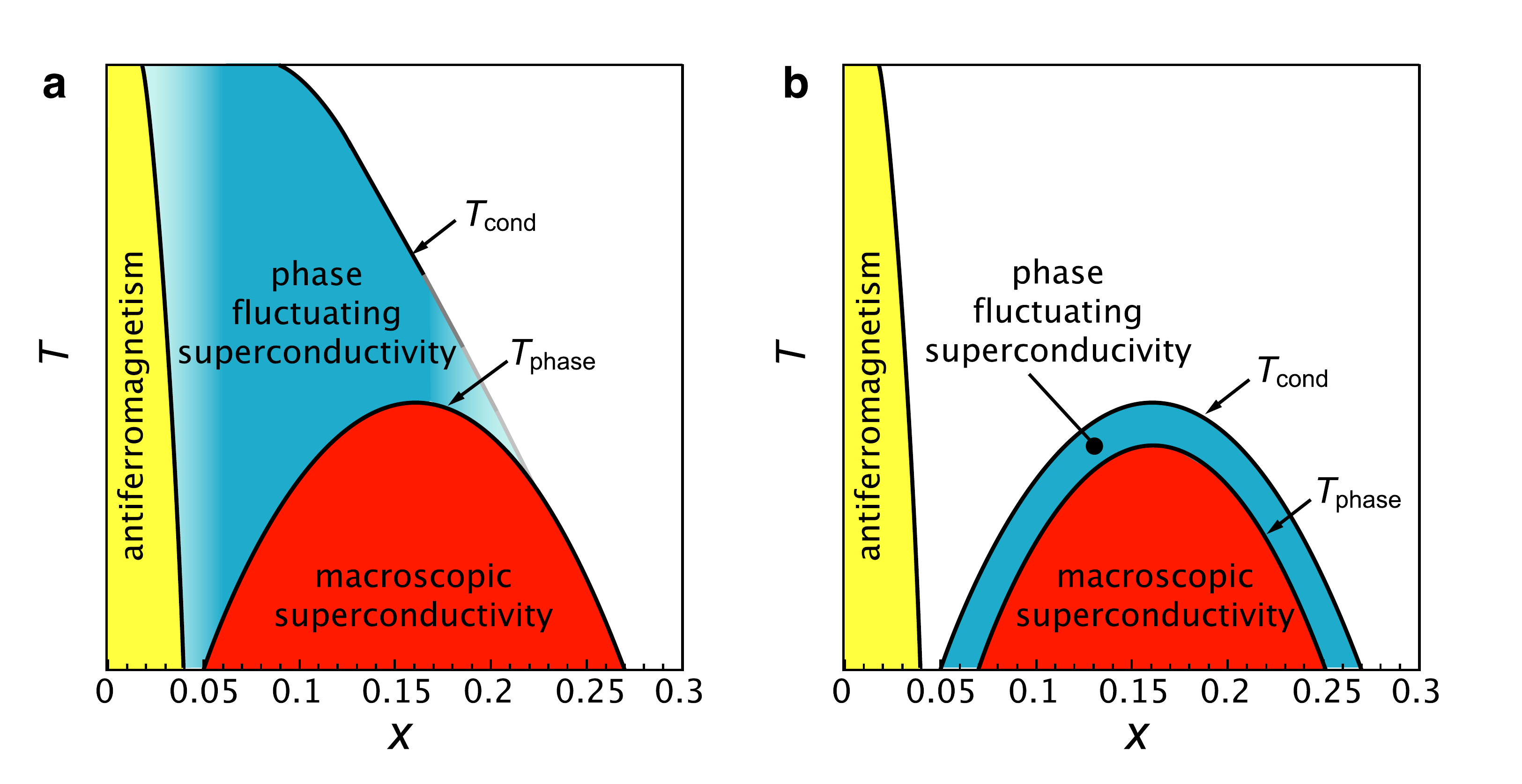}
\caption{\label{fig-uno} {\bf Two candidate scenarios for the phase diagram of the superconducting fluctuations in HTS.} In the so-called \textit{strong phase fluctuation scenario} (panel \textbf{a}) $T_{{\rm cond}}$ may be much larger than $T_{{\rm phase}}$, usually identifying $T_{{\rm cond}}$ with the pseudogap opening temperature $T_{{\rm pgap}}$\cite{Emery,\Anderson,\Lascialfari,\Ong} (several possible variations to this scenario exist, such as $T_{{\rm pgap}}$
and $T_{{\rm phase}}$ being well larger than $T_{{\rm cond}}$ only for the underdoped HTS,  or $T_{{\rm phase}}$ being identified with the Nernst onset temperature $T_{{\rm Nernst}}$, always well larger than $T_{{\rm cond}}$ but essentially constant for $x\lesssim0.16$). In panel \textbf{b}, the critical temperature shift $T_{{\rm cond}}-T_{\mathrm{phase}}$ is bound by $\lesssim5\mathrm{K}$ for all dopings; in that case $T_{{\rm pgap}}$ and $T_{{\rm Nernst}}$ would be due to electronic orders different from superconducting fluctuations (prominent examples being screened Bose pairs,\cite{GIL} charge and spin waves,\cite{\dwave} circular charge currents,\cite{\orbital,ShekhterRamshawetal2013} or oscillating charge stripes and Fermi surface reconstruction\cite{\stripes}). For simplicity, both schemas assume hole-doped HTS and omit the depressions in the superconducting temperatures often seen around $x=1/8$.}
\end{figure}

A currently common thinking about HTS is that both the superconducting transition and the anomalous features of the normal state (including also their dependences with doping) have to be understood taking into account the appearance of some new form of electron ordering below the so-called pseudogap line, $T_{{\rm pgap}}(x)$. \cite{Pines,NatPhysCoral,Franz,Tranquada} This order should have the same symmetry as the superconducting wave function and would explain the apparent reduction of the normal-state carriers' density of states occurring below $T_{{\rm pgap}}$.\cite{Pines,NatPhysCoral,Franz,Tranquada} Various different candidates of electronic states producing such order have been proposed including, \textit{e.g.}, oscillating charge stripes\cite{PRX,Cyr,KivelsonStr} possibly causing Fermi surface reconstruction\cite{PRX,Cyr}, circular orbital currents\cite{VarmaNat,ShekhterRamshawetal2013}, charge and spin waves\cite{Zhang}, or screened Bose pairs\cite{GIL}. However, probably the most popular is the one first proposed by Emery and Kivelson\cite{Emery} (and then by numerous workers, \textit{e.g.}, in\citeabajo{\phaseSinEmery}). In this last scenario, the superconducting pair condensation and the establishment of wave function phase coherence happen at different temperatures $T_{{\rm cond}}$ and $T_{{\rm phase}}$ that may be various decades of Kelvin distant from each other, especially for underdoped HTS. This situation is schematized in Fig.~\ref{fig-uno}a. The macroscopic transition temperature $T_{\rm c}$ is renormalized from
$T_{{\rm cond}}$ up to essentially $T_{{\rm phase}}$ supposedly due to strong full-critical fluctuations of the phase of the superconducting wave function as those first studied in 2D superfluids by Berezinskii, Kosterlitz and Thouless (BKT)\cite{HN}.

%%%%%%%%%%%%%%%%%%%%%%%%%%%%%%%%%%%%%%%%%%%%%%%%%%%%%%
%%%%%%%%%%%%%%%%%%%%%%%%%%%%%%%%%%%%%%%%%%%%%%%%%%%%%%

Apparent confirmations of this \textit{strong phase fluctuation scenario} for HTS were presented in a series of notable measurements of anomalously large, superconducting-like signals above $T_{\rm c}$ of the Nernst signal\cite{\OngNer}, \textit{N}, and of the magnetization\cite{\Lascialfari,\OngMag}, $M$. However, soon they were challenged by plausible alternative explanations to these observations: In the case of $M$, in terms of the unavoidable inhomogeneities due to the intrinsic random location of the non-stoichiometric dopant ions;\cite{\challengesMag} in the case of $N$, in terms of different electron orders also capable of magnetic field-dependent entropy transport, as for instance stripe
order and Fermi-surface reconstruction.\cite{\reconstruccion} These alternative explanations are coherent with a more moderate fluctuation shift of $T_{\rm c}$, with $T_{{\rm phase}}-T_{{\rm cond}}\stackrel{<}{_{\sim}}5{\rm K}$ for all dopings, as schematized in Fig.~\ref{fig-uno}b. Also, this is consistent with the scenarios in\citeabajo{\nophase,ShekhterRamshawetal2013} that identify $T_{\rm c}$ with the energy needed for the normal-state particle pairing (instead of pairing between phase-fluctuating vortices and antivortices as in the BKT-like scenario of\citeabajo{\phase}).

To shed further light over this crucial issue, here we report measurements of the\textsc{ dc} voltage-current $V-I$ curves of ${\rm La}_{2-x}{\rm Sr}_x{\rm CuO}_4$
($\mathrm{L}\mathrm{S}_{x}\mathrm{CO}$) films with different dopings. We identify a jump-like feature in the \textit{\large T}-dependence of the exponent \textit{$\mathit{\alpha}$} of \textbf{$V\propto I^{\alpha}$}, that departs in a narrow \textit{\large T}-range from the ohmic value $\alpha=1$ to a nonlinear $\alpha\geq3$. This jump, as predicted already in 1979 by Halperin and Nelson (HN),\cite{HN} is expected to occur in any BKT transition and is associated with the leap at \textbf{$T_{{\rm phase}}$} of the superfluid density (the so-called Nelson jump). Its observation is a particularly exclusive signature of the occurrence of phase coherence through a vortex-antivortex binding process. Indeed, measurements of $\alpha$ were extensively used to study the BKT transition in low-$T_{\rm c}$ superconducting films and interfaces (see, \textit{e.g.},\citeabajo{HN,HNLTSC6} and references therein). Moreover, the $\alpha$-jump is known to be almost unaffected by the existence of random superconducting inhomogeneities (as independently shown by using finite-element computations, percolation theory, renormalization group or effective-medium equations\cite{CotonHN1,CastellaniHN1,CastellaniHN2}). In spite of these advantages, the $\alpha$-jump feature has not yet been measured in HTS as a function of doping (see, \textit{e.g.},\citeabajo{HNHTSC5} and references therein), not allowing to decide between the scenarios of Fig.~\ref{fig-uno}. Getting a precise value of $T_{{\rm phase}}$ will also allow us to determine $T_{{\rm cond}}$ by analyzing the critical rounding of the ohmic resistivity above $T_{{\rm phase}}$. This rounding is due to BKT full-critical phase fluctuations immediately above $T_{{\rm phase}}$ and to Gaussian-Ginzburg-Landau (GGL)-type phase and amplitude fluctuations sufficiently above $T_{{\rm cond}}$. Importantly, to analyze the resistivity rounding so close to the transition it is necessary to take the unavoidable (intrinsic-like) chemical inhomogeneities into account, what we achieve by means of independent high-resolution magnetometry characterizations of the associated dispersion $\Delta T_{\rm c}$ of critical temperatures in each sample. 

%%%%%%%%%%%%%%%%%%%%%%%%%%%%%%%%%%%%%%%%%%%%%%%%%%%%%%
\begin{figure}[t]
\centering{}\includegraphics[width=1\columnwidth]{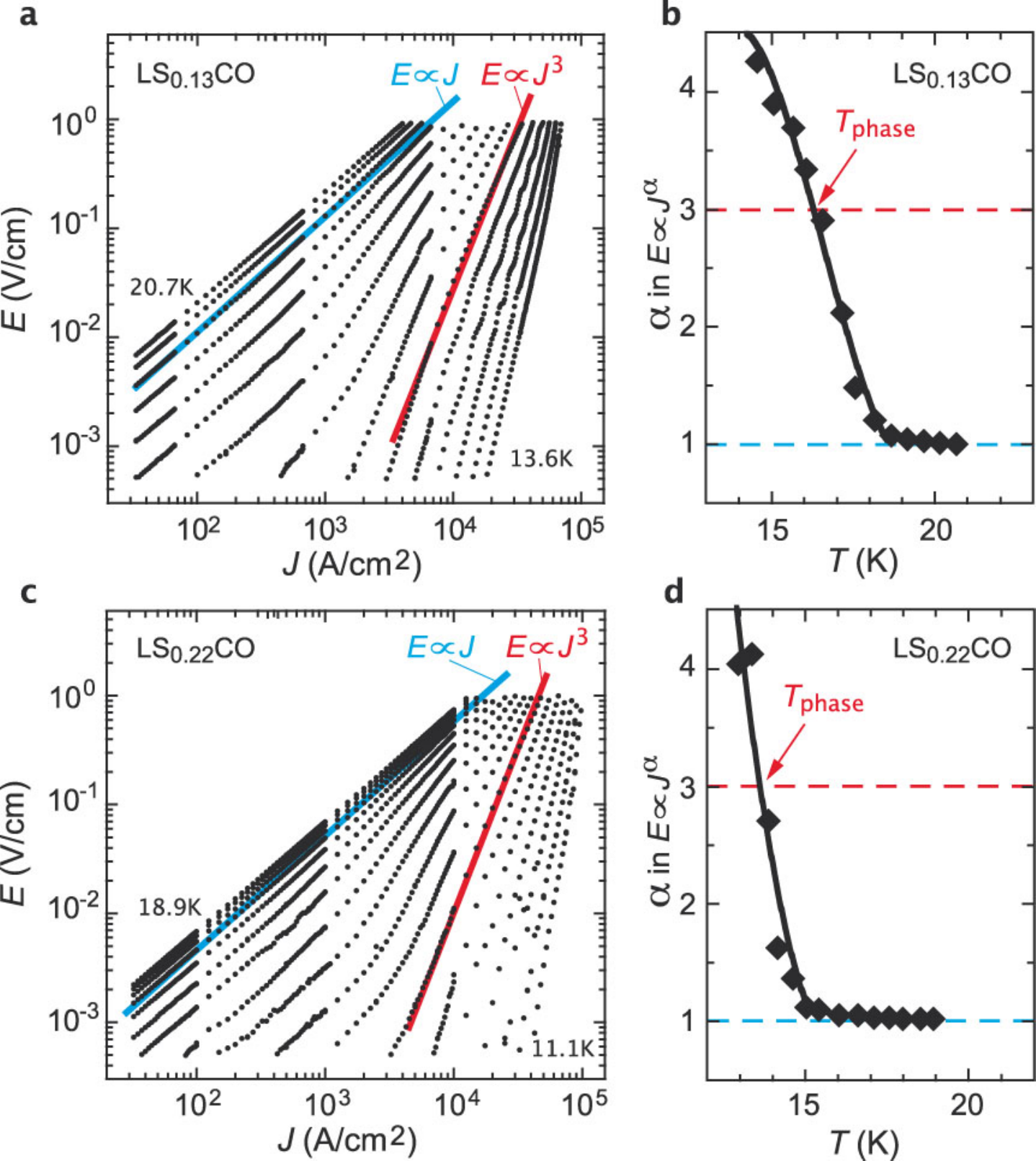}
\caption{\label{fig-dos} {\bf {\it E-J} curves and resulting $\alpha$ exponent for two example ${\rm LS}_x{\rm CO}$ films.} Panels \textbf{a} and \textbf{c} show the measured \textit{E-J} data for an underdoped and an overdoped $\mathrm{L}\mathrm{S}_{x}\mathrm{CO}$ film respectively. Panels \textbf{b} and \textbf{d} represent the corresponding $\alpha$ exponent obtained as the log-log slope of the \textit{E-J} curves in the range $10^{-3}<\mathrm{E}<0.1$ V/cm (solid lines are guides for the eyes). The temperature \textbf{$T{}_{{\rm phase}}$} corresponds to $\alpha=3$. The  results for the rest of our samples are reported as Supplementary Information.}
\end{figure}
%%%%%%%%%%%%%%%%%%%%%%%%%%%%%%%%%%%%%%%%%%%%%%%%%%%%%%

\begin{figure}[h]
\centering{}\includegraphics[width=0.78\columnwidth]{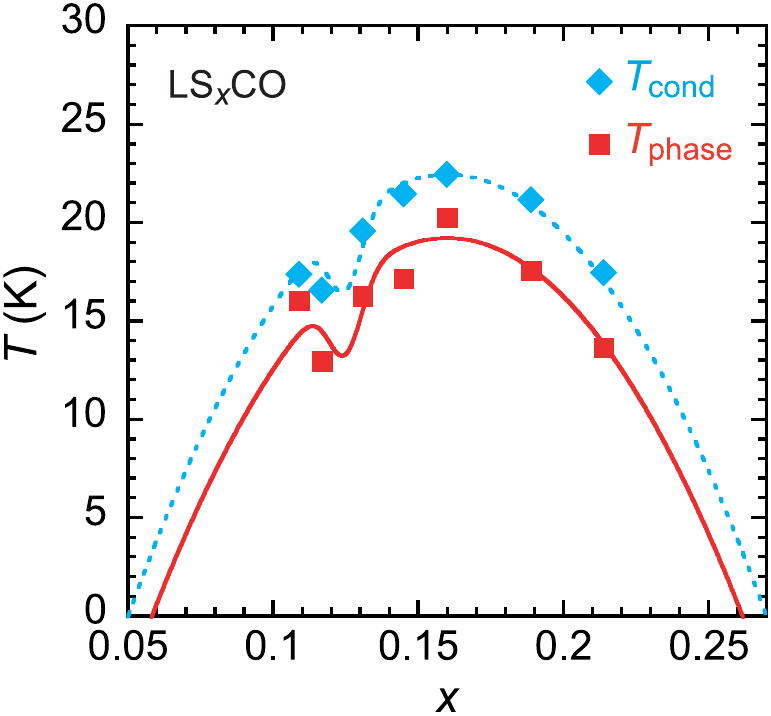}
\caption{\label{fig-cuatro} {\bf Phase diagram for $T_{\rm cond}$ and $T_{\rm phase}$ obtained from our measurements.} This figure includes the results for our entire sample set, covering doping levels $0.11\leq x\leq0.22$. Red squares correspond to \textbf{$T{}_{{\rm phase}}$}. Blue diamonds represent \textbf{$T{}_{{\rm cond}}$}. The blue dashed and red solid lines are fits using a parabolic functionality (minus a Gaussian peak centered at $x=1/8$), as $T_{\mathrm{cond}}(x)=T_{{\rm cond}}^{\mathrm{opt}}[1-((x-0.16)/0.11)^{2}]-\delta T_{\rm c 1/8}\exp[-((x-1/8)/\delta x_{1/8})^{2}]$ and $T_{{\rm phase}}(x)=T_{{\rm cond}}(x)-\Delta_{{\rm BKT}}$ (we obtain $T_{\mathrm{cond}}^{\mathrm{opt}}=22.5$K, $\delta T_{\rm c 1/8}=$ 3.6K, $\delta x_{1/8}=0.008$ and $\Delta_{{\rm BKT}}=$ 3.2K).}
\end{figure}

In panels \textbf{a} and \textbf{c} of Fig.~\ref{fig-dos} we show the electric field versus current density, $E-J$, measured for various temperatures in our films with $x=0.13$ and $x=0.22$, chosen as representative examples of underdoped and overdoped compositions (the data and analyses corresponding to the rest of our films, with $x=$ 0.11, 0.12, 0.15, 0.16 and 0.19, are documented in the Supplementary Information). The slope in these $E-J$ plots, due to their log-log scale, directly corresponds to the exponent $\alpha$. We represent the evolution with temperature of that slope in panels \textbf{b} and \textbf{d} of Fig.~\ref{fig-dos}. Noteworthy, the exponent $\alpha$ takes the ohmic value $\alpha=1$ for the higher temperatures and it abruptly departs to well larger values as the system is cooled down, with the appearance of non-ohmic characteristics happening in
a quite narrow temperature interval. According to the HN calculations for perfectly homogeneous 2D superconductors,\cite{HN} the condition $\alpha=3$ marks the temperature $T_{{\rm phase}}$; for a superconductor with a spatial distribution of $T_{\rm c}$ values, according to the calculations of\citeabajo{CotonHN1} the condition $\alpha=3$ simply marks the \textit{average} value of the phase coherence temperature (as inhomogeneities broaden somewhat the jump around the average $T_{{\rm phase}}$ but do not move the $\alpha=3$ point~\cite{CotonHN1}). We thus identify the temperature for $\alpha=3$ in our measurements with the phase coherence temperature corresponding to the nominal doping of each sample. Application of this procedure to our whole sample set, with $0.11\le x\le0.22$, leads to the $T_{{\rm phase}}(x)$ line in Fig.~\ref{fig-cuatro}. This line displays the usual dome shape regularly attributed to the transition to macroscopic superconductivity in HTS. It also includes a depression centered at $x=1/8$, often linked to stripe effects. Our measurements support then a transition accompanied by a phase-coherence vortex-antivortex binding process.

\begin{figure}[t]
\centering{}\includegraphics[width=0.88\columnwidth]{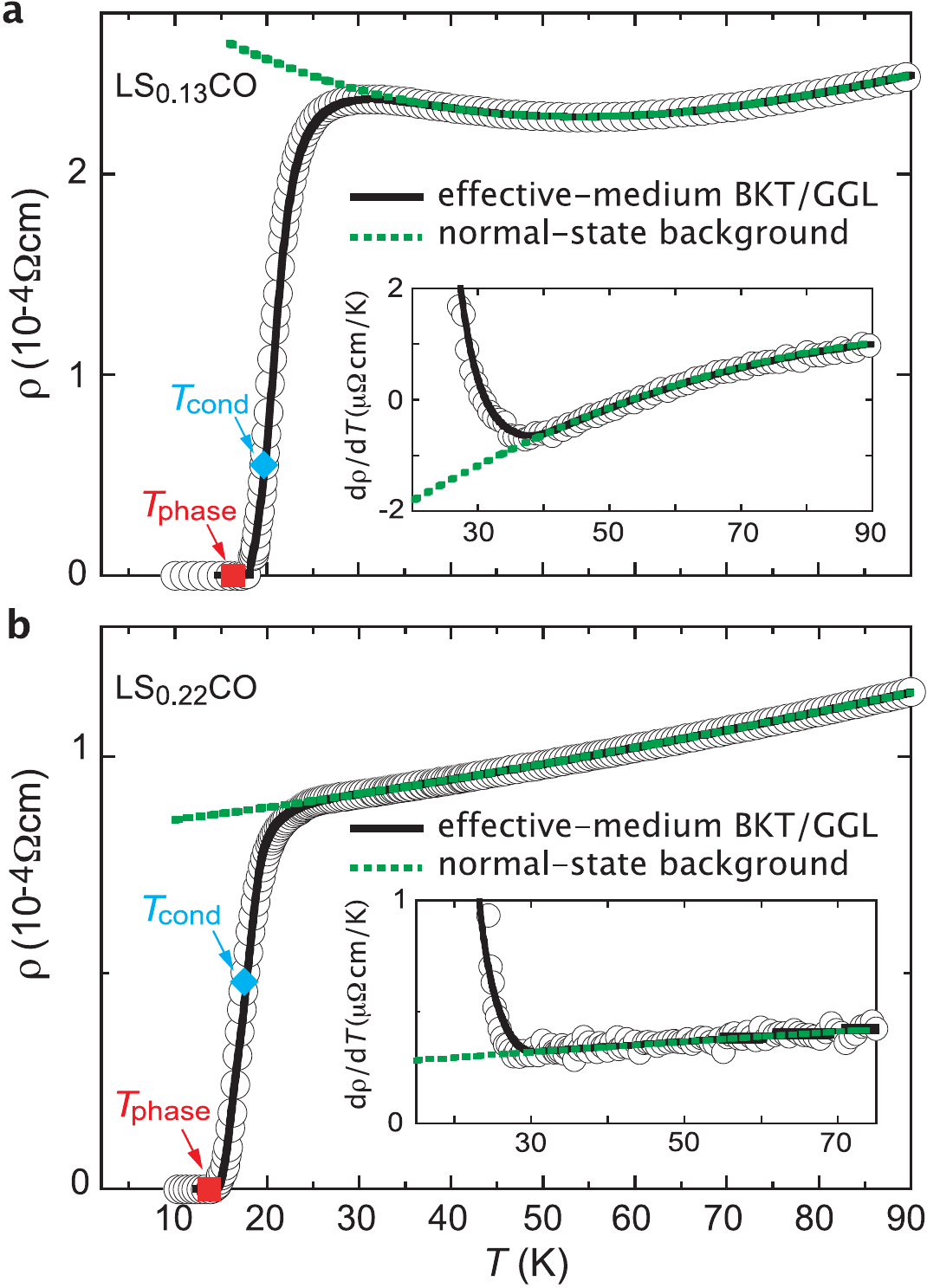}
\caption{\label{fig-tres} {\bf Low-current resistivity.} Comparison, for the same example films as in Fig.~\ref{fig-dos}, between the resistivity measurements at $J=10^{3}\mathrm{A/}\mathrm{cm}^{2}$ (open circles) and the results of the BKT/GGL theory with inhomogeneities taken into account through the effective-medium approach (black solid lines). The red squares and blue diamonds represent the \textbf{$T{}_{{\rm phase}}$} and \textbf{$T\mathrm{_{cond}}$} temperatures respectively. The green
dashed line is the normal-state background $\rho_{B}$. The insets represent the  results for $\mathrm{d}\rho/\mathrm{d}T$, focused around the onset of fluctuations corresponding to the sharp upturn of that derivative.}
\end{figure}

The precise knowledge of $T_{{\rm phase}}$ also allows to study the critical rounding of the resistivity $\rho(T)$ above the transition, extracting in the process the temperature $T_{{\rm cond}}$ for wave function condensation, and opening the opportunity to compare $T_{{\rm phase}}$ and $T_{{\rm cond}}$ in each sample. Figure~\ref{fig-tres} shows $\rho(T)$ measured with small current densities, in the same films as in\textbf{ }Fig.~\ref{fig-dos} (the results for the rest of doping levels are again documented in the Supplementary Information). Note in these $\rho(T)$ curves that the average $T_{{\rm phase}}$ (marked as a red solid square) corresponds to the tail of the resistive decay. Above it, the transition to the normal state is rounded over an easily accessible temperature range. This fluctuation rounding is commonly characterized through the so-called paraconductivity $\Delta\sigma\equiv\rho^{-1}-\rho_{B}^{-1}$ where $\rho_{B}$ is the normal-state resistivity background. Calculations are available for $\Delta\sigma$, both in terms of the BKT approach valid right above \textbf{$T{}_{{\rm phase}}$}\cite{HN} and, above $T_{{\rm cond}}$, in terms of the GGL approach\cite{AL} extended up to high-reduced temperatures\cite{CarballeiraDs,VidalEPL}. A short account of the corresponding equations is provided in the \ref{sec:eqs}, where we also describe the effective-medium formula necessary to include the effects of a possible distribution of critical temperatures. The excellent agreement between these standard equations and our $\rho(T)$ data above \textbf{$T{}_{{\rm phase}}$} is presented in Fig.~\ref{fig-tres} for the same example films as in Fig.~\ref{fig-dos}. The only free parameter in these comparisons is \textbf{$T_{\mathrm{cond}}$} (see \ref{sec:eqs}).

%%%%%%%%%%%%%%%%%%%%%%%%%%%%%%%%%%%%%%%%%%%%%%%%%%%%%%
%%%%%%%%%%%%%%%%%%%%%%%%%%%%%%%%%%%%%%%%%%%%%%%%%%%%%%

%%%%%%%%%%%%%%%%%%%%%%%%%%%%%%%%%%%%%%%%%%%%%%%%%%%%%%
%%%%%%%%%%%%%%%%%%%%%%%%%%%%%%%%%%%%%%%%%%%%%%%%%%%%%%

In Fig.~\ref{fig-cuatro} we represent as a function of the doping level \textsl{$x$} the $T_{{\rm cond}}$ and $T_{{\rm phase}}$ temperatures obtained by applying these procedures to our full sample set. Both $T_{{\rm cond}}(x)$ and $T_{{\rm phase}}(x)$ draw similar domes, displaced only a few Kelvin from each other for all dopings, the average $T_{{\rm cond}}-T_{\mathrm{phase}}$ being 3.2K. These results support then a scenario similar to the one in Fig.~\ref{fig-uno}b, in which the vortex-antivortex fluctuations and bindings shift only moderately the critical temperatures, and being the normal-state quasiparticle pairing energies the ones that primarily dominate the values and $x$-dependence of the transition temperatures.

In conclusion, our findings indicate that macroscopic superconductivity occurs in our films via a BKT-like vortex-antivortex binding, with the superconducting wave function involved in that process having a $T_{\rm cond}$  located only a few Kelvin above $T_{\rm phase}$. This last result is at odds with the strong phase fluctuation scenario\cite{Franz,Emery,OngNeruno,Lascialfari2002,AndersonNat,OngMagPRB} but is instead compatible  with  different forms of electronic order proposed to explain the pseudogap and superconducting phases\cite{PRX,Cyr,KivelsonStr,VarmaNat,ShekhterRamshawetal2013,Zhang,GIL}.  These orders also include forms of local electronic correlations not necessarily producing a superconducting wave function (due, {\it e.g.},  to screening by stripe boundaries\cite{KivelsonStr,PRX,Cyr} or by other electrons\cite{GIL}). We note also that, while the jump-like onset of the $E-J$ exponent $\alpha\geq3$ is quite exclusive  of a  BKT-like transition, we cannot   discard that below this jump $\alpha$ could be  affected by additional contributions due to some of these coexisting forms of order. For instance, charge density waves\cite{Cyr} are known to produce nonlinear contributions below the transition, with a relatively small typical exponent of about\cite{Fisher} 1.5, and if they emerge associated to vortex cores in cuprates (as recently proposed in\citeabajo{Wu}) they  could  also add to the $E-J$ response below the BKT jump.

\medskip{}

\footnotesize

\textbf{Methods}

\textit{Samples} used for our measurements are LS$_{x}$CO thin films with dopings $0.11\leq x\leq0.22$, grown over (100)$\mathrm{SrTi}\mathrm{O}_{3}$ substrates by using a procedure specifically aimed at improving the superconducting homogeneity.\cite{CotonFilms} As also described in\citeabajo{CotonFilms}, the critical temperature dispersion, $\Delta T_{\mathrm{c}}$, of each film was measured through high-precision measurements of the zero-field-cooled magnetic susceptibility $\chi$: Because $\left|\chi\right|$ in the fully superconducting state is orders of magnitude larger than in the normal state (even in presence of superconducting fluctuations), ${\rm d}\chi/{\rm d}T$ is proportional to the $T_{\mathrm{c}}$ distribution in the film. We thus obtain $\Delta T_{\mathrm{c}}$ from the width of the ${\rm d}\chi/{\rm d}T$ peak at the transition (see also the Supplementary Information for a summary of the quantitative results). The obtained $\Delta T_{\mathrm{c}}$ are among the lowest ever reported for the LS$_{x}$CO family (including bulk and single crystals) and agree with the predictions of the models for intrinsic-disorder inhomogeneities with a residual constant dispersion $\Delta T\mathrm{_{c}^{res}}$ of only 0.5K.

\textit{Transport measurements} were done in a 4-probe in-line configuration over microbridges lithographed in the films, and using current pulses of about 1~ms to avoid self-heating effects.

\textit{The normal-state background }$\rho_{B}$ was obtained as follows: First, we noted that the resistivity data present a clear change of tendency in their slope occurring at a well-defined temperature $T_{{\rm upturn}}$ well above the transition (as clearly visible in the insets of Fig.~\ref{fig-tres}). This change of behaviour signals the first visible deviations from the purely normal-state behaviour. Therefore, we obtain $\rho_{B}$ as a simple fit to the $\rho$ and ${\rm d}\rho/{\rm d}T$ data above such $T_{\mathrm{upturn}}$. For all dopings, we fitted a 50K-wide region and used a quadratic functionality for ${\rm d}\rho_{B}/{\rm d}T$. Note that this procedure is an improvement with respect to most previous studies of $\Delta\sigma$ in HTS, that had to cope with less definite criterions for the temperature well above the transition above which the normal-state behaviour could be fitted.

%%%%%%%%%%%%%%%%%%%%%%%%%%%%%%%%%%%%%%%%%%%%%%%%%%%%%%
%%%%%%%%%%%%%%%%%%%%%%%%%%%%%%%%%%%%%%%%%%%%%%%%%%%%%%
%%%%%%%%%%%%%%%%%%%%%%%%%%%%%%%%%%%%%%%%%%%%%%%%%%%%%%
%%%%%%%%%%%%%%%%%%%%%%%%%%%%%%%%%%%%%%%%%%%%%%%%%%%%%%

%%%%%%%%%%%%%%%%%%%%%%%%%%%%%%%%%%%%%%%%%%%%%%%%%%%%%%
%%%%%%%%%%%%%%%%%%%%%%%%%%%%%%%%%%%%%%%%%%%%%%%%%%%%%%
%%%%%%%%%%%%%%%%%%%%%%%%%%%%%%%%%%%%%%%%%%%%%%%%%%%%%%
%%%%%%%%%%%%%%%%%%%%%%%%%%%%%%%%%%%%%%%%%%%%%%%%%%%%%%

\setlength{\parindent}{0cm}

\vskip3em\textbf{Acknowledgements} 

We acknowledge fruitful conversations about the general implications of our results with  F.\,Lalibert\'e, A.J.\,Leggett,  D.\,Pavuna and L.\,Taillefer, about electron pairing located into stripes with J.M.\,Tranquada, and about the  nonlinear response of charge density waves below $T_{\rm phase}$ with J.B.\,Goodenough. We also thank J.\,Mosqueira and B.\,Mercey for their support for  the samples' preparation and measurements. We acknowledge funding by Spain's Ministerio de Ciencia e Innovaci\'on project FIS2010-19807 and by Xunta de Galicia's projects 2010/XA043 and 10TMT206012PR. All projects are co-funded by ERDF from the European Union. N.C.\,acknowledges financial support from Spain's Ministerio de Econom\'{\i}a y Competitividad under project FIS2007-63709 (MEC-FEDER) trough a FPI grant.

%\textbf{Author contributions}

\vskip3em{\sl This Letter is accompanied by a Supplementary Information addenda. N.C., M.V.R.\,and F.V.\,contributed equally to this work. The authors declare no competing financial interests. }
%
%N.C.~performed the experiments. N.C.~and M.V.R.~analyzed the data. M.V.R.~and F.V.~conceived the experiment and the leading scientific questions. All authors contributed to the interpretation of the data and the writing of the manuscript.

%\textbf{Additional information}

%\textbf{Competing financial interests}

\newpage{}

%%%%%%%%%%%%%%%%%%%%%%%%%%%%%%%%%%%%%%%%%%%%%%%%%%%%%%
%%%%%%%%%%%%%%%%%%%%%%%%%%%%%%%%%%%%%%%%%%%%%%%%%%%%%%
%%%%%%%%%%     END OF MAIN PAPER     %%%%%%%%%%%
%%%%%%%%%%%%%%%%%%%%%%%%%%%%%%%%%%%%%%%%%%%%%%%%%%%%%%
%%%%%%%%%%%%%%%%%%%%%%%%%%%%%%%%%%%%%%%%%%%%%%%%%%%%%%

\clearpage{}
%%%%%%%%%%%%%%%%%%%%%%%%%%%%%%%%%%%%%%%%%%%%%%%%%%%%%%
%%%%%%%%%%%%%%%%%%%%%%%%%%%%%%%%%%%%%%%%%%%%%%%%%%%%%%
%%%%%%%%%%%%%%%%     SUPPLEMENTARY INFORMATION     %%%%%%%%%%%%%%%%%
%%%%%%%%%%%%%%%%%%%%%%%%%%%%%%%%%%%%%%%%%%%%%%%%%%%%%%
%%%%%%%%%%%%%%%%%%%%%%%%%%%%%%%%%%%%%%%%%%%%%%%%%%%%%%

%%%%%%%%%%%%%%%%     initial formatting and title page of supplementary information    %%%%%%%%%%%%%%%%%

\begin{center}
\onecolumngrid
\par\end{center}

\mbox{}\vspace{2cm}

\begin{center}
\textsf{\textbf{\LARGE Supplementary Information for the work:}}
\par\end{center}{\LARGE \par}

\medskip{}

\begin{center}
\textbf{\textsl{\LARGE ``Critical temperatures for superconducting\\ phase-coherence and condensation in La$_{2-x}$Sr$_{x}$CuO$_{4}$''}}
\par\end{center}{\LARGE \par}

\vspace{2cm}

\begin{center}
\textsf{\textbf{\Large Noelia Cot\'on, Manuel V. Ramallo, F\'elix Vidal}}
\par\end{center}{\Large \par}

\vspace{2cm}

\noindent \begin{center}
\textsl{\large LBTS, Departamento de F\'{\i}sica da Materia Condensada, \\ Universidade de Santiago de Compostela, ES-15782 Spain.}
\par\end{center}{\large \par}

\titleformat{\section} 
{\normalfont\fontsize{13pt}{13pt}\selectfont\sf\bfseries }{\thesection:}{1em}{}

\renewcommand{\thesection}{\mbox{Supplementary Section~\arabic{section}}}
% redefine the command that creates the section no. 
\renewcommand{\theequation}{\arabic{equation}}
% redefine the command that creates the equation no. 
\setcounter{equation}{0}  
% reset equation counter 
\renewcommand{\thefigure}{\arabic{figure}}
\renewcommand{\figurename}{\mbox{\sf Supplementary Figure}}
% redefine the command that creates the figure no. 
\setcounter{figure}{0}  
% reset figure counter 
\renewcommand{\thetable}{\arabic{table}}
\renewcommand{\tablename}{\mbox{\sf Supplementary Table}}
% redefine the command that creates the table no.
\setcounter{table}{0}  
% reset table counter 
\renewcommand{\thepage}{s\arabic{page}}
% redefine the command that creates the page no.
\setcounter{page}{1}  
% reset page counter 
\mbox{}\vspace{3cm}\mbox{}
\renewcommand{\normalsize}{\fontsize{10pt}{10pt}\selectfont}\normalsize \sf
\setlength{\baselineskip}{14pt}
\setlength{\parskip}{10pt}
\renewcommand{\large}{\fontsize{12pt}{12pt}\selectfont}
\renewcommand{\small}{\fontsize{9pt}{9pt}\selectfont}

%%%%%%%%%%%%%%%%%%%%%%%%%%%%%%%%%%%%%%%%%%%%%%%%%%%%%%
%%%%%%%%%%%%%%%%%%%%%%%%%%%%%%%%%%%%%%%%%%%%%%%%%%%%%%
%%%%%%%%%%%%%%%%%%%%%%%%%%%%%%%%%%%%%%%%%%%%%%%%%%%%%%
%%%%%%%%%%%%%%%%%%%%%%%%%%%%%%%%%%%%%%%%%%%%%%%%%%%%%%

\clearpage{}\newpage{}\hrule

\section{\label{sec:eqs}\ }

\noindent \textbf{\large Summary of the theoretical predictions for the paraconductivity $\Delta\sigma$\vspace{2cm}
}{\large \par}

We abridge here the existing theoretical predictions for $\Delta\sigma$ in the in-plane direction of a layered superconductor in the 2D limit, in the Berezinskii-Kosterlitz-Thouless (BKT) and Gaussian-Ginzburg-Landau (GGL) regimes of the superconducting fluctuations. These formulas are used in our present work to analyze $\rho(T)$ above $T_{\mathrm{phase}}$ with the aim of extracting $T_{\mathrm{cond}}$. The effects on $\Delta\sigma$ of critical-temperature inhomogeneities (unavoidable in $\mathrm{L}\mathrm{S}_{x}\mathrm{CO}$ due to the non-stoichiometric nature of its doping) are also discussed, including a summary of the corresponding effective-medium approach.
Finally, we comment on the main constraints affecting the values of the parameters involved in these formulas. In what follows, $d$ is the distance between adjacent CuO$_{2}$ planes, $\hbar$ is the reduced Planck's constant, $k_{B}$ is the Boltzmann's constant and $e$ is the electron charge.

%%%%%%%%%%%%%%%%%%%%%%%%%%%%%%%%%%%%%%%%%%%%%%%%%%%%%%
\mbox{}\vskip0.5em\mbox{}\textbf{A. $\Delta\sigma$ in the temperature region of BKT superconducting fluctuations}

Immediately above $T_{{\rm phase}}$ (in the often-called strong phase fluctuation or full-critical regime) the relevant superconducting excitations are topological (vortex and antivortex positions) and they may be well described by the renormalization group approaches.\cite{HN} In that regime, the coherence length $\xi$ depends exponentially on $\left(b_{0}\Delta_{{\rm BKT}}/T_{\mathrm{phase}}\right)^{1/2}t^{-1/2}$, where $\Delta_{{\rm BKT}}=T_{{\rm cond}}-T_{{\rm phase}}$ is the BKT displacement, $t\equiv(T-T_{\mathrm{phase}})/T_{\mathrm{phase}}$ is the BKT reduced-temperature and $b_{0}$ is a constant of the order of unity. \cite{HN} As first predicted by HN, from the Josephson's relation it follows then a fluctuation conductivity given by:\cite{HN}
%%%%%%%%%%%%%%%%%%%%%%%%%%%%%%%%%%%%%%%%%%%%%%%%%%%%%%
\begin{equation}
\Delta\sigma_{{\rm BKT}}=A_{{\rm BKT}}\exp\sqrt{\frac{4b_{0}\Delta_{\mathrm{BKT}}}{T_{\mathrm{phase}}}\frac{1}{t}},\label{eq:BKT}
\end{equation}
%%%%%%%%%%%%%%%%%%%%%%%%%%%%%%%%%%%%%%%%%%%%%%%%%%%%%%
where the constant $A_{{\rm BKT}}$ may be directly obtained by continuity of $\Delta\sigma_{{\rm BKT}}$ with the results of the contiguous GGL temperature region described in the next subsection.

%%%%%%%%%%%%%%%%%%%%%%%%%%%%%%%%%%%%%%%%%%%%%%%%%%%%%%
\mbox{}\vskip1em\mbox{}\textbf{B. $\Delta\sigma$ in the temperature region of GGL superconducting fluctuations}

Sufficiently above $T_{{\rm cond}}$ (in the often-called amplitude fluctuation or GGL regime) topological excitations will be no longer dominant, and the fluctuations must become small Gaussian perturbations of the amplitude and phase of the superconducting wave function, given by the mean-field-like Ginzburg-Landau approach. In this conventional
regime $\xi\varpropto\varepsilon^{-1/2}$ with $\varepsilon$ being the GGL reduced-temperature $\varepsilon\equiv\ln(T/T_{{\rm cond}})$ ($\simeq(T-T_{\mathrm{cond}})/T_{{\rm cond}}$ for small $\varepsilon$). According to the classical calculations by Aslamazov and Larkin (AL)\cite{AL} the fluctuation conductivity is then $(e^{2}/16\hbar d)\varepsilon^{-1}$. However, the latter result is not expected to remain valid in the high reduced-temperature region $\varepsilon\stackrel{>}{_{\sim}}0.1$, as it does not take into account the quantum limits to the short-wavelength fluctuations.\cite{VidalEPL} Full expressions accounting for the short-wavelength effects were calculated in\citeabajo{CarballeiraDs} on the grounds of a total-energy cutoff approach,\cite{VidalEPL} obtaining
%%%%%%%%%%%%%%%%%%%%%%%%%%%%%%%%%%%%%%%%%%%%%%%%%%%%%%
\begin{equation}
\Delta\sigma_{{\rm GGL}}=\frac{e^{2}}{16\hbar d}\frac{1}{\varepsilon}\left(1-\frac{\varepsilon}{\varepsilon^{\mathrm{c}}}\right)^{2}.\label{eq:GGL}
\end{equation}
%%%%%%%%%%%%%%%%%%%%%%%%%%%%%%%%%%%%%%%%%%%%%%%%%%%%%%
In this equation, $\varepsilon^{\mathrm{c}}$ is a cutoff reduced-temperature near which fluctuations experience a rapid fall, so that for $\varepsilon>\varepsilon^{\mathrm{c}}$
they effectively become null. Therefore, in our case a first crude approximation would be $\varepsilon^{\mathrm{c}}\simeq\ln(T_{{\rm upturn}}/T_{{\rm cond}}$).\renewcommand{\thefootnote}{\fnsymbol{footnote}}\footnotemark[2]\footnotetext[2]{\small\sf Alternatively, it is also possible to estimate\cite{VidalEPL} $\varepsilon^{\rm c}\simeq 0.55$ on the grounds of simple BCS-like arguments that ultimately reflect that superconducting fluctuations will not occur at the temperatures in which $\xi$ is already below its $T=0$K value $\xi_{T=0\rm K}$, that is the minimum superconducting volume that the uncertainty principle allows; as shown in detail in our Supplementary Section 3B, within our experimental uncertainties both estimates for the precise value of $\varepsilon^{\rm c}$ will be equally compatible with the measurements in the present work.}

Calculating the precise boundary temperature between the BKT and the GGL regions is, in full rigour, still an open theoretical problem. However, the Levanyuk-Ginzburg criterion for the breakdown of the Gaussian hypothesis of the GGL approach is commonly accepted as a valid first crude approximation for that boundary. This predicts for layered superconductors in the 2D limit\cite{RamalloeLG} a GGL breakdown above $T_{{\rm cond}}$ at about $T_{{\rm LG}}\equiv T_{{\rm cond}}\exp(\varepsilon_{\mathrm{LG}})$
with $\varepsilon_{{\rm LG}}\equiv k_{{\rm B}}/(4\pi\xi^{2}(0)d\Delta C)$, where $\xi(0)$ is the GGL amplitude of the coherence length and $\Delta C$ is the heat capacity mean-field jump per unit volume. Estimates for HTS suggest that\cite{RamalloeLG} $\varepsilon_{{\rm LG}}\sim10^{-2}$.

%%%%%%%%%%%%%%%%%%%%%%%%%%%%%%%%%%%%%%%%%%%%%%%%%%%%%%
\mbox{}\vskip1em\mbox{}\textbf{C. Effects of critical temperature inhomogeneities}

When trying to study in terms of the above predictions the $\rho(T)$ rounding in actual superconductors, it is important to take critical temperature inhomogeneities into account, particularly when analyzing the data so close to the transition as in the present work. It is important to emphasize that, due to the non-stoichiometry of $\mathrm{L}\mathrm{S}_{x}\mathrm{CO}$, even for ideal samples grown under perfectly uniform conditions those inhomogeneities will exist. In our case, the full-width at half-maximum (FWHM), $\Delta T_{\mathrm{c}}$, of the critical temperatures distribution peak was measured in each of our samples by means of high-precision magnetometry (as described in the Methods section of our main text, and in further detail in\citeabajo{CotonFilms}). These measurements lead to $\Delta T_{\mathrm{c}}$ values in our films among the lowest ever reported using magnetometry for their corresponding dopings in the $\mathrm{L}\mathrm{S}_{x}\mathrm{CO}$ family (including bulk and single crystals). This is illustrated in the Supplementary Fig.~\ref{Figura-good}, where the $\Delta T_{\mathrm{c}}$ obtained in previous works by other authors using similar magnetometry characterizations (Refs.\citeabajo{MosqueiraPtos} to\citeabajo{Marin}) are compared with ours. In this figure, the $T_{\mathrm{c}}$ used to normalize the results corresponds to the maximum of the $\mathrm{d}\chi/\mathrm{d}T$ peak of the transition. This figure also plots the $\Delta T_{\mathrm{c}}/T_{\mathrm{c}}$ prediction resulting from a model with intrinsic-disorder inhomogeneities with a constant residual dispersion, $\Delta T{}_{\mathrm{c}}^{\mathrm{res}}$ (see\citeabajo{CotonFilms} for a complete calculation). The resulting $\Delta T{}_{\mathrm{c}}^{\mathrm{res}}=0.5$K is also a particularly small value.\cite{CotonFilms} The numerical $\Delta T_{\mathrm{c}}$ values for each of our films are  listed in the Supplementary Table~\ref{Tabla1}.

%%%%%%%%%%%%%%%%%%%%%%%%%%%%%%%%%%%%%%%%%%%%%%%%%%%%%%
\begin{figure}
\begin{centering}
\includegraphics[width=0.45\columnwidth]{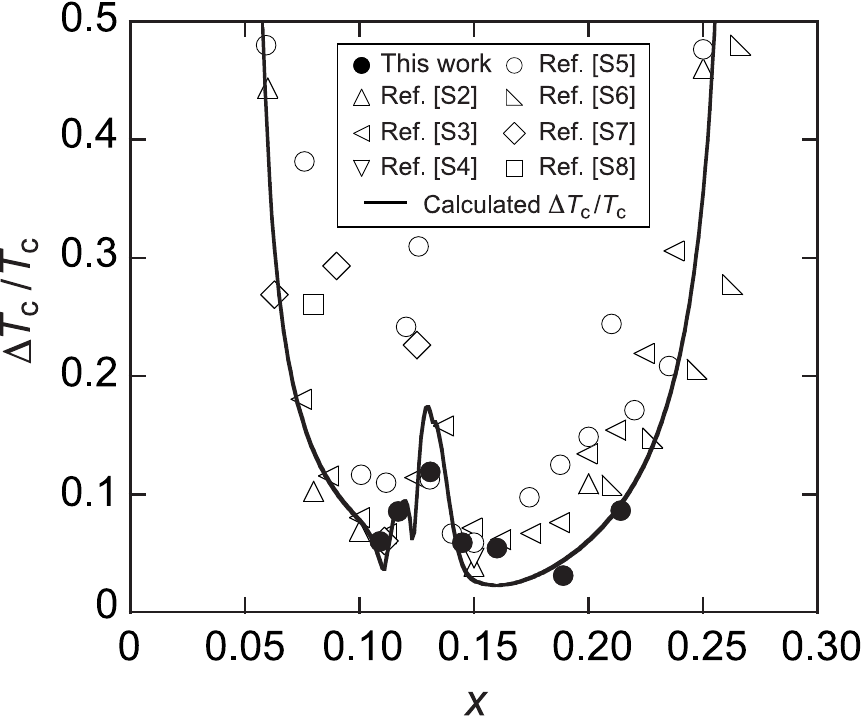}
\par\end{centering}
\caption{\label{Figura-good}\small\sf Critical temperature dispersion as a function of the doping level $x$ in $\mathrm{L}\mathrm{S}_{x}\mathrm{CO}$ for the samples studied in this work (solid symbols) and for other works (open symbols) reporting some of the narrowest transitions obtained in the $\mathrm{L}\mathrm{S}_{x}\mathrm{CO}$ family using a magnetic susceptibility $\chi$ analysis comparable to ours. $\Delta T_{\mathrm{c}}$ corresponds to FWHM of the $\mathbf{\mathrm{d}\chi/\mathrm{d}}T$ peak measured under zero-field-cooled conditions, and $T_{\mathrm{c}}$ to the maximum of that peak. The solid line is the $\Delta T_{\mathrm{c}}/T_{\mathrm{c}}$ calculated by taking into account the intrinsic disorder of the dopant locations and a residual dispersion of $\Delta T{}_{\mathrm{c}}^{\mathrm{res}}=$0.5K (see\citeabajo{CotonFilms} for details on the calculation). This figure illustrates both the quality of our films and, more importantly, that even for the best $\mathrm{L}\mathrm{S}_{x}\mathrm{CO}$ the inhomogeneities due to the intrinsic doping disorder cannot be avoided.}
\end{figure}
%%%%%%%%%%%%%%%%%%%%%%%%%%%%%%%%%%%%%%%%%%%%%%%%%%%%%%

The effects of these critical temperature inhomogeneities over $\Delta\sigma$ may be easily taken into account by means of the effective-medium approximation for superconductors with a random normal distribution of critical temperatures uniformly distributed in space:\cite{MazaEM,CotonHN2}
%%%%%%%%%%%%%%%%%%%%%%%%%%%%%%%%%%%%%%%%%%%%%%%%%%%%%%
\begin{equation}
\int_{0}^{\infty}\frac{\sigma_{T'_{{\rm cond}}}-\sigma}{\sigma_{T'_{{\rm cond}}}+2\sigma}\exp\left[-\left(\frac{T'_{{\rm cond}}-T_{{\rm cond}}}{\Delta T_{\rm c}/(2\sqrt{\ln2})}\right)^{2}\right]\frac{\mathrm{d}T'_{{\rm cond}}}{\Delta T_{\mathrm{c}}}=0.\label{eq:effmed}
\end{equation}
%%%%%%%%%%%%%%%%%%%%%%%%%%%%%%%%%%%%%%%%%%%%%%%%%%%%%%
Here $\sigma$ is the global conductivity of the sample, $\sigma_{T'_{{\rm cond}}}$ is the conductivity of a domain having a single value $T'_{{\rm cond}}$ of the condensation temperature, $T_{{\rm cond}}$ is the average of the condensation temperature distribution in the sample, and $\Delta T_{\rm c}$ is the FWHM dispersion of that distribution. We  always consider $T_{\mathrm{cond}}-T\mathrm{_{phase}}$ as uniform in each superconductor. Although equation (\ref{eq:effmed}) is not explicit for $\sigma$, the set of equations (\ref{eq:BKT}) to (\ref{eq:effmed}) may be numerically solved by modern computers with ease.

Note that we have limited the measurements in our present work to a doping range in which the inhomogeneities remain relatively small,  $\Delta T_c/T_c\stackrel{<}{_\sim}0.1$, so to keep down as much as feasible  the influence of these inhomogeneity effects.

%%%%%%%%%%%%%%%%%%%%%%%%%%%%%%%%%%%%%%%%%%%%%%%%%%%%%%
\mbox{}\vskip1em\mbox{}\textbf{D. Constraints for the theory parameters}

It is important to realize, when comparing the above equations (\ref{eq:BKT}) to (\ref{eq:effmed}) to the actual data in HTS, that the values of most of the involved parameters are constrained by conditions that significantly limit the degrees of freedom in those comparisons. It is useful to summarize here these constraints:

First of all, and as emphasized in our main text, $T_{{\rm phase}}$ is the temperature at which $\alpha=3$, and $\Delta T_{\rm c}$ may also be independently measured by means of magnetometry as the FWHM of the $\mathrm{d}\chi/\mathrm{d}T$ peak at the transition (the latter has been done for our films in\citeabajo{CotonFilms}, see also Supplementary Table~\ref{Tabla1}).

For the parameter $\varepsilon^{\mathrm{c}}$, \textit{i.e.}, the reduced-temperature above which there are no superconducting fluctuations, in our analyses we will use the constraint $\varepsilon^{\mathrm{c}}=\ln(T_{{\rm upturn}}/T_{{\rm cond}})$ where $T_{\mathrm{upturn}}$ is the temperature where a clear change in $\mathrm{d}\rho/\mathrm{d}T$ behaviour is observed in our films. For the Levanyuk-Ginzburg temperature we apply an allowance $0.5\times10^{-2}\leq\varepsilon_{\mathrm{LG}}\leq2\times10^{-2}$,
coherently with the estimates in\citeabajo{RamalloeLG}. Regarding $b_{0}$, it is expected to be a constant of the order of unity in the BKT predictions; we therefore impose in our comparisons the constraint that \textit{all} of the seven $\rho(T)$ data fits corresponding to our seven doping levels share a common value for $b_{0}$. This will lead to $b_{0}=4$.\renewcommand{\thefootnote}{\fnsymbol{footnote}}\footnotemark[2]\footnotetext[2]{\small\sf We checked that for the parameters mentioned in this paragraph ($\varepsilon^{\rm c}$, $\varepsilon_{\rm LG}$ and $b_0$) removing these constraints in our comparisons with the $\rho(T)$ data may affect the quality of the fits, but do not significantly affect the results for $T_{\rm cond}$, nor $T_{\rm phase}$, which are the two main concerns of this paper (this aspect is discussed in detail in the Supplementary Sections 3A and 3B).}

The only fully free parameter remaining in the comparison of equations (\ref{eq:BKT}) to (\ref{eq:effmed}) with the $\rho(T)$ data in each of our samples is, therefore, $T_{{\rm cond}}$.

%%%%%%%%%%%%%%%%%%%%%%%%%%%%%%%%%%%%%%%%%%%%%%%%%%%%%%
%%%%%%%%%%%%%%%%%%%%%%%%%%%%%%%%%%%%%%%%%%%%%%%%%%%%%%
%%%%%%%%%%%%%%%%%%%%%%%%%%%%%%%%%%%%%%%%%%%%%%%%%%%%%%
%%%%%%%%%%%%%%%%%%%%%%%%%%%%%%%%%%%%%%%%%%%%%%%%%%%%%%

\clearpage{}\newpage{}\hrule

%%%%%%%%%%%%%%%%%%%%%%%%%%%%%%%%%%%%%%%%%%%%%%%%%%%%%%
\section{\label{sec:results}\ }

\noindent \textbf{\large Results for $x=$0.12, 0.15, 0.16 and 0.19\vspace{1.2cm}}{\large \par}

In our main text, the Figs.~\ref{fig-dos} and~\ref{fig-tres} illustrate our $E-J$ and $\rho$ measurements and analyses for only two of our samples (those with dopings $x=$0.13 and 0.19). Here we report the results of applying the same procedures of measurement and analysis to the rest of our sample set.

%%%%%%%%%%%%%%%%%%%%%%%%%%%%%%%%%%%%%%%%%%%%%%%%%%%%%%
\textbf{A. Underdoped compositions $x<0.16$}

\begin{figure}[h]
\begin{centering}
\includegraphics[width=0.7\columnwidth]{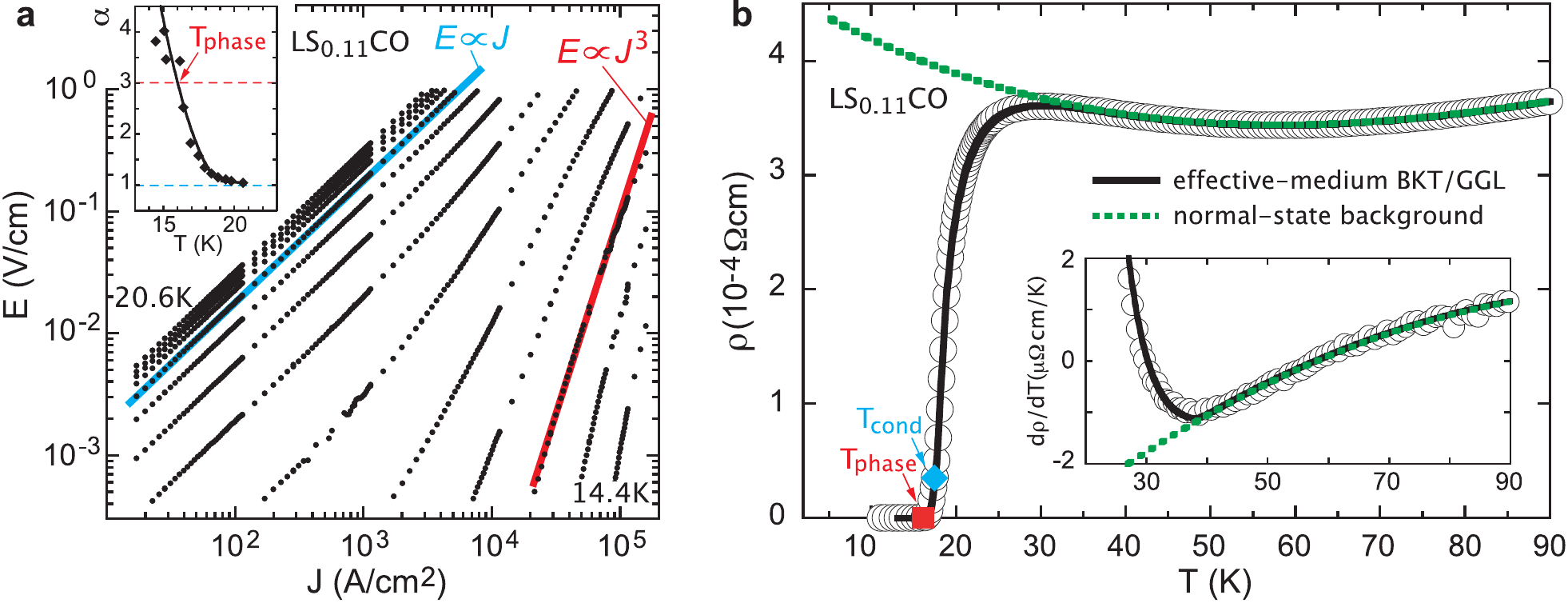}
\par\end{centering}
\caption{\label{Figura-EJ_rho-LSCO_0p11} \small\sf Results for the $\mathrm{L}\mathrm{S}_{0.11}\mathrm{CO}$ sample of our measurements and data analyses of (\textbf{a}) the\textit{ $E-J$} curves  and  (\textbf{b}) the resistivity $\rho$ at low current densities $J=10^{3}\mathrm{A/}\mathrm{cm}^{2}$. The exponent $\alpha$ (inset of panel \textbf{a}) was obtained as the log-log slope of the \textit{$E-J$} curves for $10^{-3}<\mathit{\mathrm{E}}<0.1$ V/cm. We also indicate $T_{\mathrm{phase}}$ and $T_{\mathrm{cond}}$, obtained respectively from the $\alpha=3$ condition and from the comparison of $\rho(T)$ with equations (\ref{eq:BKT}) to (\ref{eq:effmed}). This comparison corresponds to the parameter values in the Supplementary Table~\ref{Tabla1} and to the solid black line in panel \textbf{b} and its inset (the green dashed line corresponds to the normal-state background).}
\end{figure}

\begin{figure}[H]
\begin{centering}
\includegraphics[width=0.7\columnwidth]{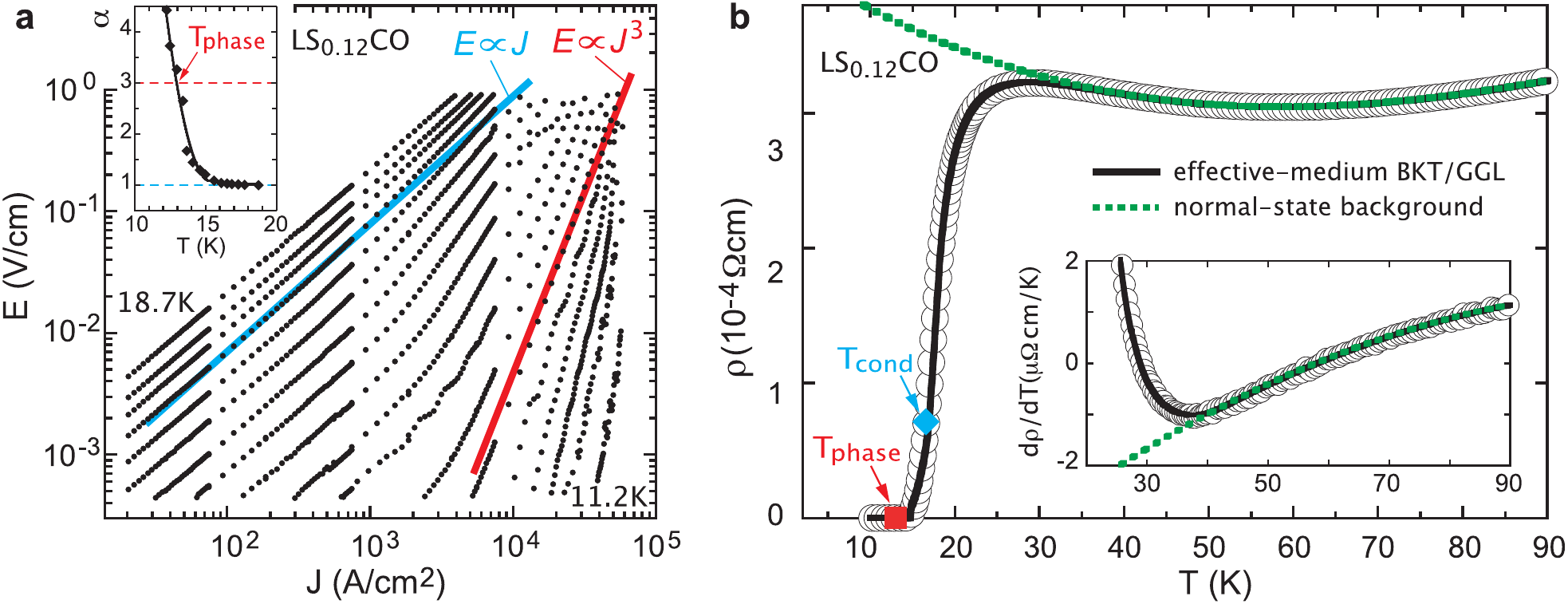}
\par\end{centering}
\caption{\label{Figura-EJ_rho-LSCO_0p12} \small\sf Results for the $\mathrm{L}\mathrm{S}_{0.12}\mathrm{CO}$ sample of our electric transport measurements and data analyses, including the same features as described in detail in the caption of the Supplementary Fig.~\ref{Figura-EJ_rho-LSCO_0p11}.}
\end{figure}

\begin{figure}[H]
\begin{centering}
\includegraphics[width=0.7\columnwidth]{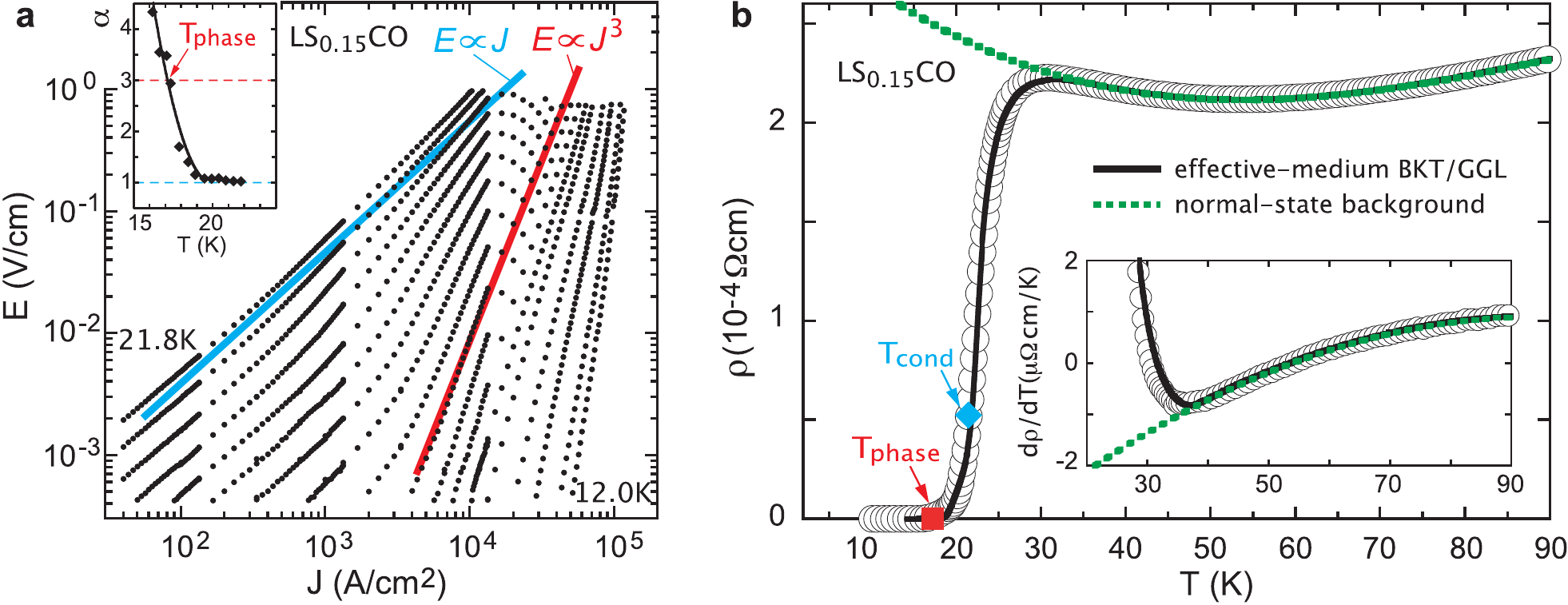}
\par\end{centering}
\caption{\label{Figura-EJ_rho-LSCO_0p15} \small\sf Results for the $\mathrm{L}\mathrm{S}_{0.15}\mathrm{CO}$ sample of our electric transport measurements and data analyses, including the same features as described in detail in the caption of the Supplementary Fig.~\ref{Figura-EJ_rho-LSCO_0p11}.}
\end{figure}

The Supplementary Figs.~\ref{Figura-EJ_rho-LSCO_0p11} to~\ref{Figura-EJ_rho-LSCO_0p15} show our \textit{$E-J$ }and $\rho$ measurements as a function of temperature, for our films with dopings $x=$0.11, 0.12 and 0.15. They also illustrate the main features of our data analyses, performed as described in the main text of our article and in the \ref{sec:eqs} (including also the parameter constraints mentioned in the Supplementary Subsection 1A). These analyses lead to  parameter values given in the Supplementary Table~\ref{Tabla1}.

%%%%%%%%%%%%%%%%%%%%%%%%%%%%%%%%%%%%%%%%%%%%%%%%%%%%%%
%%%%%%%%%%%%%%%%%%%%%%%%%%%%%%%%%%%%%%%%%%%%%%%%%%%%%%

%%%%%%%%%%%%%%%%%%%%%%%%%%%%%%%%%%%%%%%%%%%%%%%%%%%%%%
%%%%%%%%%%%%%%%%%%%%%%%%%%%%%%%%%%%%%%%%%%%%%%%%%%%%%%

%%%%%%%%%%%%%%%%%%%%%%%%%%%%%%%%%%%%%%%%%%%%%%%%%%%%%%
%%%%%%%%%%%%%%%%%%%%%%%%%%%%%%%%%%%%%%%%%%%%%%%%%%%%%%

%%%%%%%%%%%%%%%%%%%%%%%%%%%%%%%%%%%%%%%%%%%%%%%%%%%%%%
\textbf{B. Optimally-doped composition $x=0.16$}

\begin{figure}[H]
\begin{centering}
\includegraphics[width=0.7\columnwidth]{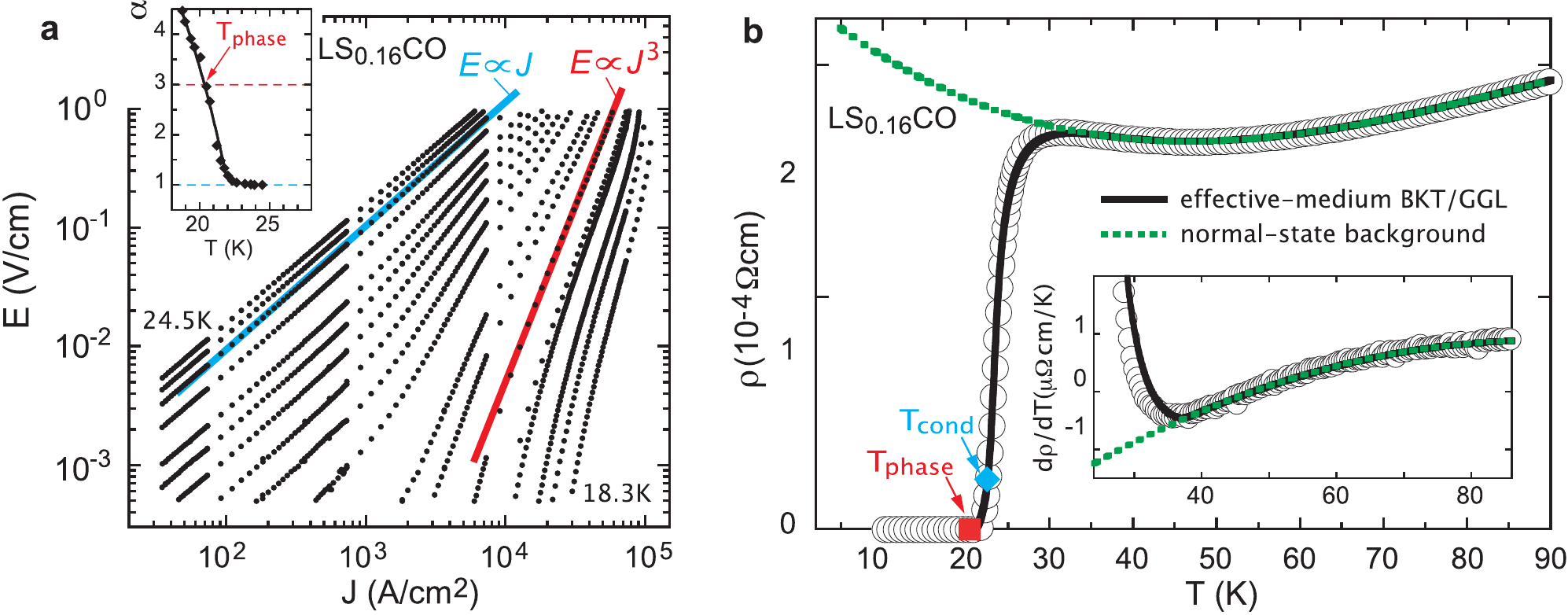}
\par\end{centering}
\caption{\label{Figura-EJ_rho-LSCO_0p16} \small\sf Results for the $\mathrm{L}\mathrm{S}_{0.16}\mathrm{CO}$ sample of our electric transport measurements and data analyses, including the same features as described in detail in the caption of the Supplementary Fig.~\ref{Figura-EJ_rho-LSCO_0p11}.}
\end{figure}

The Supplementary Fig.~\ref{Figura-EJ_rho-LSCO_0p16} shows the \textit{$E-J$} and $\rho$ measurements for our optimally-doped film $x=$0.16, and their analyses performed as in the previous subsection for the underdoped samples. The Supplementary Table~\ref{Tabla1}  includes the parameter values corresponding to these results.

%%%%%%%%%%%%%%%%%%%%%%%%%%%%%%%%%%%%%%%%%%%%%%%%%%%%%%
%%%%%%%%%%%%%%%%%%%%%%%%%%%%%%%%%%%%%%%%%%%%%%%%%%%%%%

%%%%%%%%%%%%%%%%%%%%%%%%%%%%%%%%%%%%%%%%%%%%%%%%%%%%%%
\textbf{C. Overdoped compositions $x>0.16$}

\begin{figure}[H]
\begin{centering}
\includegraphics[width=0.7\columnwidth]{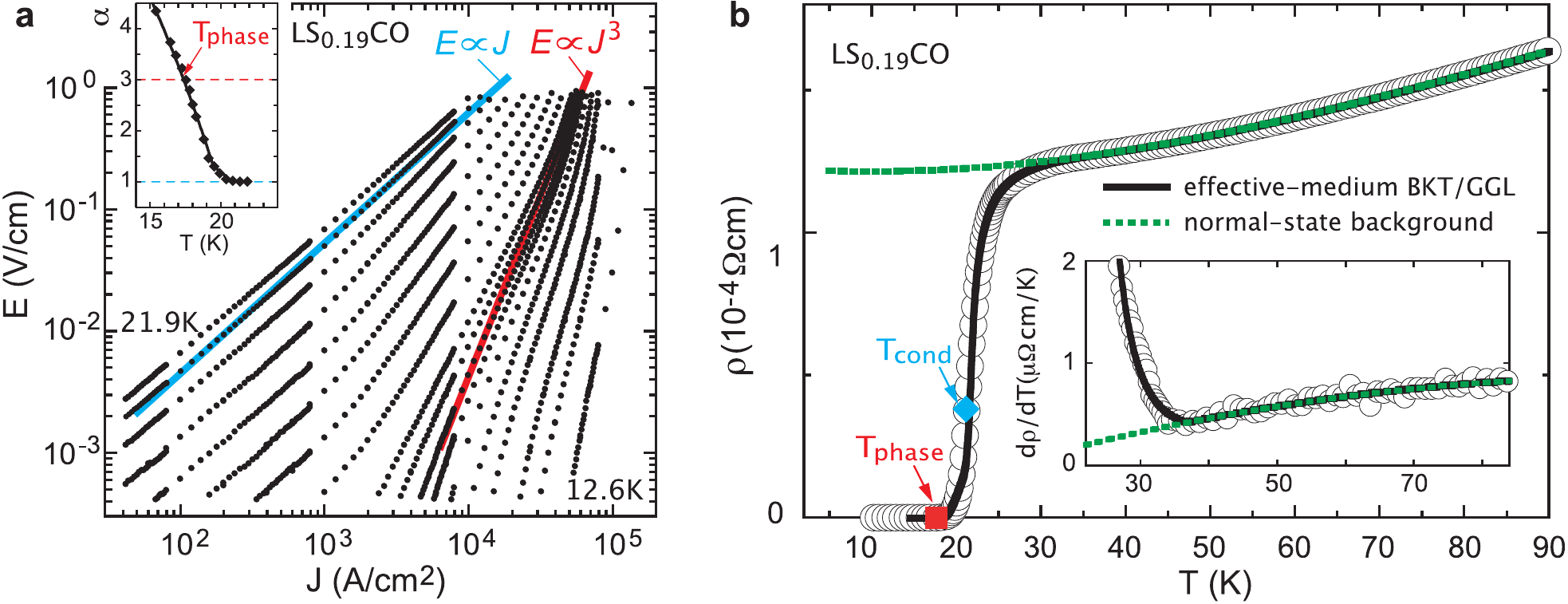}
\par\end{centering}
\caption{\label{Figura-EJ_rho-LSCO_0p19} \small\sf Results for the $\mathrm{L}\mathrm{S}_{0.19}\mathrm{CO}$ sample of our electric transport measurements and data analyses, including the same features as described in detail in the caption of the Supplementary Fig.~\ref{Figura-EJ_rho-LSCO_0p11}.}
\end{figure}

The Supplementary Fig.~\ref{Figura-EJ_rho-LSCO_0p19} shows the \textit{$E-J$} and $\rho$ measurements for the overdoped composition $x=$0.19 (please find in our main text the corresponding results for the also overdoped $x=$0.22). This figure also illustrates the main features of the analyses performed over these data employing the same procedures as described in the previous subsections for the underdoped and optimally-doped compositions. The Supplementary Table~\ref{Tabla1} includes the parameter values corresponding to these results.

%%%%%%%%%%%%%%%%%%%%%%%%%%%%%%%%%%%%%%%%%%%%%%%%%%%%%%
%%%%%%%%%%%%%%%%%%%%%%%%%%%%%%%%%%%%%%%%%%%%%%%%%%%%%%

%%%%%%%%%%%%%%%%%%%%%%%%%%%%%%%%%%%%%%%%%%%%%%%%%%%%%%
\begin{table}[H]
\mbox{}\vspace{3cm}\\ 
\begin{centering}
\begin{tabular}{|c|c|c|c|c|c|c|}
\hline 
sample & \mbox{}\hspace{1em}\mbox{}$T_{{\rm phase}}\;({\rm K})$\mbox{}\hspace{1em}\mbox{} & \mbox{}\hspace{1.6em}\mbox{}$T_{{\rm cond}}\;({\rm K})$\mbox{}\hspace{1.6em}\mbox{} & \mbox{}\hspace{1.6em}\mbox{}$\Delta T_{\mathrm{c}}\;({\rm K})$\mbox{}\hspace{1.6em}\mbox{} & \mbox{}\hspace{1em}\mbox{}$\varepsilon^{\mathrm{c}}=\ln(T_{\mathrm{upturn}}/T_{\mathrm{cond}})$\mbox{}\hspace{1em}\mbox{} & \mbox{}\hspace{2em}\mbox{}$\varepsilon_{\mathrm{LG}}$\mbox{}\hspace{2em}\mbox{} & \mbox{}\hspace{2em}\mbox{}$b_{0}$\mbox{}\hspace{2em}\mbox{}\tabularnewline
\hline 
\hline 
\mbox{}\hspace{2em}\mbox{}LS\textsubscript{0.11}CO\mbox{}\hspace{2em}\mbox{} & \mbox{}\hspace{2em}\mbox{}16.1\mbox{}\hspace{2em}\mbox{} & \mbox{}\hspace{2em}\mbox{}17.4\mbox{}\hspace{2em}\mbox{} & \mbox{}\hspace{2em}\mbox{}1.1\mbox{}\hspace{2em}\mbox{} & 0.8 & 0.010 & 4\tabularnewline
\hline 
LS\textsubscript{0.12}CO & 13.0 & 16.6 & 1.4 & 0.9 & 0.015 & 4\tabularnewline
\hline 
LS\textsubscript{0.13}CO & 16.3 & 19.6 & 2.3 & 0.7 & 0.020 & 4\tabularnewline
\hline 
LS\textsubscript{0.15}CO & 17.2 & 21.5 & 1.3 & 0.5 & 0.010 & 4\tabularnewline
\hline 
LS\textsubscript{0.16}CO & 20.3 & 22.5 & 1.3 & 0.5 & 0.005 & 4\tabularnewline
\hline 
LS\textsubscript{0.19}CO & 17.6 & 21.2 & 0.7 & 0.5 & 0.005 & 4\tabularnewline
\hline 
LS\textsubscript{0.22}CO & 13.7 & 17.5 & 1.6 & 0.5 & 0.020 & 4\tabularnewline
\hline 
\end{tabular}
\par\end{centering}
\caption{\label{Tabla1} \small\sf  Main parameters for our $\mathrm{L}\mathrm{S}_{x}\mathrm{CO}$ films. $T_{\mathrm{phase}}$ and $T_{\mathrm{cond}}$ are obtained respectively from the $\alpha=3$ condition and the comparison with the effective-medium BKT/GGL approach (equations (\ref{eq:BKT}) to (\ref{eq:effmed})). $\Delta T_{\mathrm{c}}$ is obtained as the FWHM of the $\mathbf{\mathrm{d}\chi/\mathrm{d}}T$ transition peak as reported in\citeabajo{CotonFilms}. $T_{\mathrm{upturn}}$ (and hence $\varepsilon^{\mathrm{c}}$) corresponds to the location of the sharp departure from the parabolic dependence in the $\mathbf{\mathrm{d}\rho/\mathrm{d}}T$ versus \textit{$T$}
curves. The parameters $\varepsilon_{\mathrm{LG}}$ and $b_{0}$ correspond to the best-fit values for equations (\ref{eq:BKT}) to (\ref{eq:effmed}), proceeding in these fits always with the constraints that $0.5\times10^{-2}\leq\varepsilon_{\mathrm{LG}}\leq2\times10^{-2}$ (in agreement with the estimates in\citeabajo{RamalloeLG}) and that $b_{0}$ takes the same value for all of the studied samples. As shown in \ref{sec:AdditionalConsiderations}A, relaxing these constraints to use other values of $\varepsilon_{\mathrm{LG}}$ and $b_{0}$ affects the quality of the fits but not significantly the values of $T_{\mathrm{cond}}$, nor $T_{\mathrm{phase}}$, that are the main concerns of this paper. Also using $\varepsilon^{\mathrm{c}}$ different from $\ln(T_{\mathrm{upturn}}/T_{\mathrm{cond}})$ does not affect the $T_{\mathrm{cond}}$ and $T_{\mathrm{phase}}$ obtained from these theory comparisons (see \ref{sec:AdditionalConsiderations}B). Other parameters for these films (including their detailed structural characterization) have been reported in\citeabajo{CotonFilms}.}
\end{table}
%%%%%%%%%%%%%%%%%%%%%%%%%%%%%%%%%%%%%%%%%%%%%%%%%%%%%%

%%%%%%%%%%%%%%%%%%%%%%%%%%%%%%%%%%%%%%%%%%%%%%%%%%%%%%
%%%%%%%%%%%%%%%%%%%%%%%%%%%%%%%%%%%%%%%%%%%%%%%%%%%%%%
%%%%%%%%%%%%%%%%%%%%%%%%%%%%%%%%%%%%%%%%%%%%%%%%%%%%%%
%%%%%%%%%%%%%%%%%%%%%%%%%%%%%%%%%%%%%%%%%%%%%%%%%%%%%%

\clearpage{}\newpage{}\hrule

%%%%%%%%%%%%%%%%%%%%%%%%%%%%%%%%%%%%%%%%%%%%%%%%%%%%%%
\section{\label{sec:AdditionalConsiderations}\ }

\noindent \textbf{\large Additional considerations on the $\Delta\sigma$ analyses\vspace{2cm}
}{\large \par}

\textbf{A. \ Analyses imposing the BKT-like behaviour well above $T_{90\%}$}

In our main text, and in the rest of sections of this Supplementary Information, we have analyzed our $\rho(T)$ data imposing a value of $\varepsilon_{\mathrm{LG}}\sim10^{-2}$. This produces very good agreement with the data. It also produces values of $T_{\mathrm{cond}}$ in the lower half of the $\rho(T)$ transition: In particular, those $T_{\mathrm{cond}}$ are always below $T_{90\%}$, with $T_{90\%}$ defined by the relation $\rho(T_{90\%})=0.9\rho_{B}(T_{90\%})$ (a simple way to signal the upper part of the macroscopic resistance transition). In fact, the results of these analyses lead to a region of dominance of phase fluctuations (or BKT-like region, $T\lesssim T_{\mathrm{cond}}\exp(\varepsilon_{\mathrm{LG}})$) entirely located well below $T_{90\%}$. However, in the ``strong phase fluctuation scenario'' advocated for cuprates by many authors,\cite{Franz,Emery,AndersonNat,Lascialfari2002,OngNeruno,OngMagPRB} the BKT-like region must include temperatures well above the main drop of the $\rho(T)$ transition, \textit{i.e.}, well above $T_{90\%}$.

Therefore, it may be interesting to explore what happens if we analyze our data imposing a region of strong phase fluctuations covering temperatures above $T_{90\%}$. For that, we have to compare the $\rho(T)$ data at $T\gtrsim T_{90\%}$ with the BKT prediction for the paraconductivity, equation (\ref{eq:BKT}). For increased generality, we do not constraint our analyses with any requirement about a crossover to a GGL regime at larger temperatures; this also implies to take the constant $A{}_{\mathrm{BKT}}$ of equation (\ref{eq:BKT}) as a fully free constant. Noteworthy, equation (\ref{eq:BKT}) leads to upwards concavity of the $\Delta\sigma(t)$ dependence (in log-log representation, and for all parameter values). This already indicates that this equation will not be able to explain the data at too high temperatures because, as it may be seen in the Supplementary Fig.~\ref{Figura-paraconductividad}, the experimental $\Delta\sigma(t)$ displays at those temperatures a downturn of significant \textit{downwards} concavity.

%%%%%%%%%%%%%%%%%%%%%%%%%%%%%%%%%%%%%%%%%%%%%%%%%%%%%%
\begin{figure}
\begin{centering}
\includegraphics[width=0.7\columnwidth]{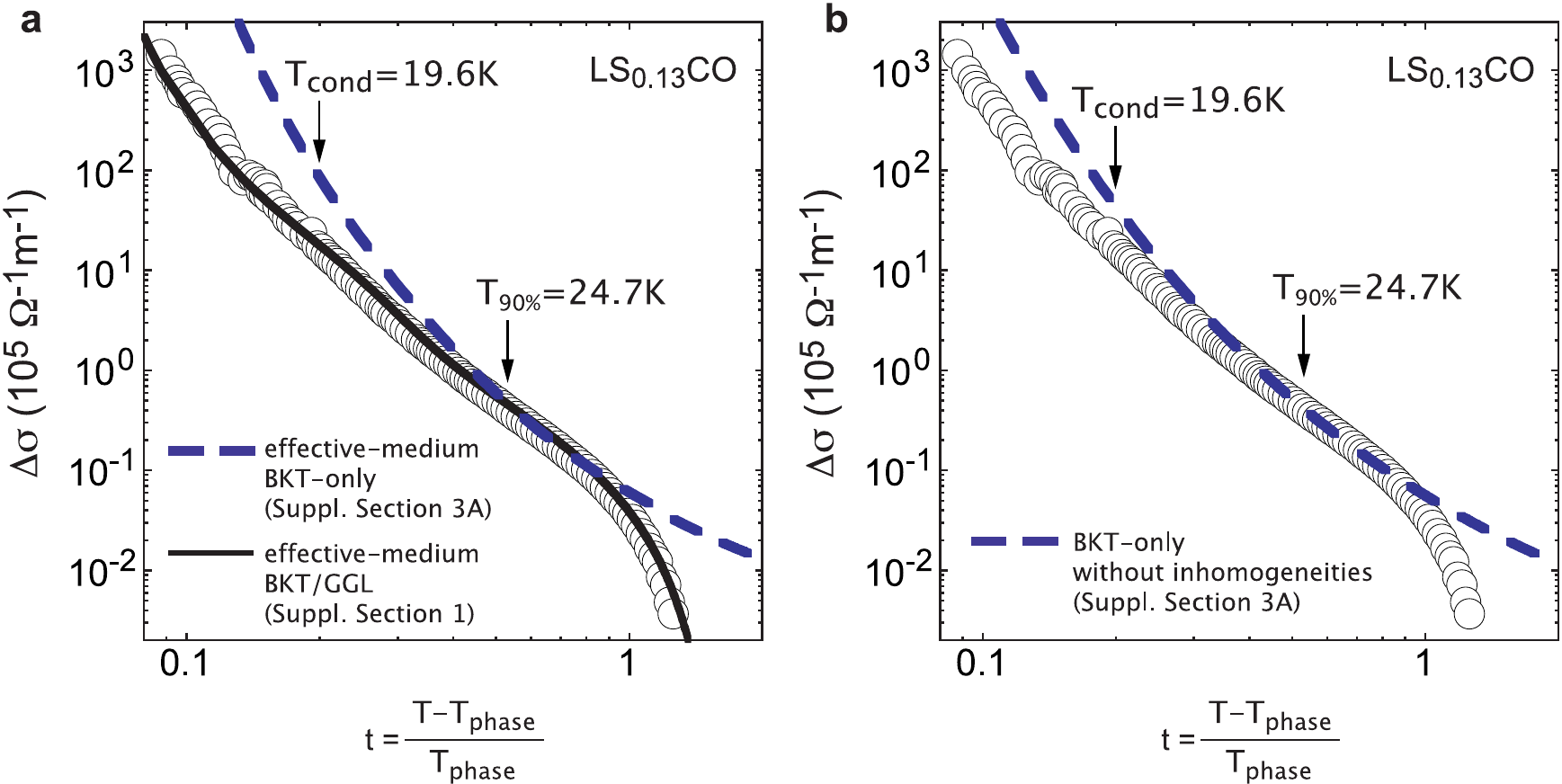}
\par\end{centering}
\caption{\label{Figura-paraconductividad} \small\sf  Paraconductivity of our $\mathrm{L}\mathrm{S}_{0.13}\mathrm{CO}$ film versus the BKT reduced-temperature $t$.\textbf{ }Open circles show the measured paraconductivity. The dashed lines in \textbf{a} and \textbf{b} are fits of the BKT theory imposing it to be valid right above the temperature $T_{90\%}$. In panel \textbf{a}, the effects of critical-temperature inhomogeneities are included in the fit by means of the effective-medium approximation (\textit{i.e.}, the fit corresponds to equations (\ref{eq:BKT}) and (\ref{eq:effmed})). In panel \textbf{b}, the inhomogeneities are not taken into account (the fit corresponds to equation (\ref{eq:BKT}) alone). For completeness, panel \textbf{a} also shows, as a solid line, the $\Delta\sigma(T)$ fit that results from the fluctuation scenario proposed in the main text (\textit{i.e.}, the effective-medium BKT/GGL approach, or equations (\ref{eq:BKT}), (\ref{eq:GGL}) and (\ref{eq:effmed})); the temperature $T_{\mathrm{cond}}$ marked as a reference in both panels is the one obtained with the latter approach.}
\end{figure}
%%%%%%%%%%%%%%%%%%%%%%%%%%%%%%%%%%%%%%%%%%%%%%%%%%%%%%

We also show in the Supplementary Fig.~\ref{Figura-paraconductividad} a fit using equation (\ref{eq:BKT}) to our paraconductivity data in the temperature range extending from $T_{90\%}$ up to the temperature where the experimental $\Delta\sigma(t)$ displays a downturn of significant downwards concavity. Given that the value of $T_{\mathrm{phase}}$ is fixed by the $\alpha=3$ condition in the \textit{$V-I$} measurements, the two free parameters in these comparisons are the product $b_{0}\Delta_{\mathrm{BKT}}$ and the amplitude $A{}_{\mathrm{BKT}}$. The resulting fit is plotted as a dashed line in the Supplementary Fig.~\ref{Figura-paraconductividad}. This figure shows the illustrating case of sample $\mathrm{L}\mathrm{S}_{0.13}\mathrm{CO}$; in the rest of our samples, we obtained fits of either comparable or even worse quality. Also, this figure uses a log-log representation of  $\Delta\sigma$ versus a reduced-temperature, a type of plot frequently used in papers by experts in paraconductivity; the fit proposed in our main analyses (solid line in this figure) provides an excellent data agreement even in this representation. In the panel \textbf{a} of the Supplementary Fig.~\ref{Figura-paraconductividad} the effects of critical temperature inhomogeneities have been taken into account using the same procedure as in our main text (\textit{i.e.}, using the effective-medium equation (\ref{eq:effmed}) and the $\Delta T_{\mathrm{c}}$ value resulting from the SQUID measurements). For completeness, we also show in the panel \textbf{b} of that figure the results by using equation (\ref{eq:BKT}) with no inhomogeneities taken into account.

All these results evidence that the approach with a BKT-like region above $T_{90\%}$ explored in this section always leads to a significantly poorer agreement with the data than the one proposed in our main text: The upwards concavity of the theory is clearly excessive for both the low- and high-\textit{$t$} regions (for the high-\textit{$t$} region, even the sign of the concavity is wrong). In fact, the theory can be seen as merely a tangent to the experiments.

It may be also investigated whether the evident disagreement between the BKT-only approach and the data for high $t$ may be fixed by imposing a crossover to a GGL-type behaviour, locating now the crossover at those high temperatures (this, in fact, amounts to repeating our main analyses but simply with different $\varepsilon_{\rm LG}$ and $b_0$ parameter values). In that case, the corresponding theory line in Supplementary Fig.~\ref{Figura-paraconductividad}a would just jump at the crossover temperature from the dashed to the solid line. The low-$t$ disagreement would remain. Even in that scenario the solid BKT/GGL line corresponds to the value of $T_{\rm cond}$ given in Fig.~\ref{fig-tres}a of our main text, \textit{i.e.}, it is always $T_{\rm cond}<T_{90\%}$. In other words, the shift of the BKT/GGL crossover temperature, $T_{\rm cond}{\rm exp}(\varepsilon_{\rm LG})$, would happen via an increase of the Levanyuk-Ginzburg reduced-temperature $\varepsilon_{\rm LG}$ well above $10^{-2}$, and not through an increase of $T_{\rm cond}$. Therefore, for $T_{\rm cond}$, which is the main issue of interest for us in these analyses, the results of our main text (Figs.~\ref{fig-tres} and \ref{fig-cuatro}) holds irrespectively of these variations in the analysis.

\mbox{}

%%%%%%%%%%%%%%%%%%%%%%%%%%%%%%%%%%%%%%%%%%%%%%%%%%%%%%
\textbf{B. \ Analysis of the uncertainty in the value of $\varepsilon^{\mathrm{c}}$}

In our main text we have analyzed our $\rho(T)$ data always considering the parameter $\varepsilon^{\mathrm{c}}$ constrained to be $\mathrm{\ln}(T_{\mathrm{upturn}}/T_{\mathrm{cond}})$, consistently with the interpretation in the extended GGL approach of $\varepsilon^{\mathrm{c}}$ as the reduced-temperature above which fluctuations disappear in a homogeneous sample.\cite{CarballeiraDs,VidalEPL} This led to $\varepsilon^{\mathrm{c}}$ in the range $0.5\leq\varepsilon^{\mathrm{c}}\leq0.9$ (see Supplementary Table~\ref{Tabla1}) with only three samples having $\varepsilon^{\mathrm{c}}\text{>0.6}$. These values are close to the theoretical prediction $\varepsilon^{\mathrm{c}}=0.55$ obtained in\citeabajo{VidalEPL} using the BCS approach in the clean limit (or $\varepsilon^{\mathrm{c}}=0.6$ in the dirty limit). However,
it may be asked what reasons may exist for the $\varepsilon^{\mathrm{c}}$ variability.

Let us show here that the data in all our films are in fact compatible with fits imposing $\varepsilon^{\mathrm{c}}=0.55$ if we consider for the inhomogeneities the possibility of distributions of critical temperatures with slight asymmetry (instead of the simple and symmetrical Gaussian distributions used in our main analyses). The explored asymmetry
takes the form of an additional upper tail comprising just about between 2\% and 4.5\% of the sample, and is too small to be resolved even by our SQUID measurements of the $T_{\mathrm{c}}$ distributions. We will also show that this $\varepsilon^{\mathrm{c}}$ uncertainty does not affect the values of the average $T_{\mathrm{phase}}$ or $T_{\mathrm{cond}}$ obtained in our analyses for each film. We illustrate these analyses in the Supplementary Fig.~\ref{Figura-Asymmetric_Gauss-LSCO_0p12} with the sample $\mathrm{LS{}_{0.12}CO}$ (that is the one with larger $\varepsilon^{\mathrm{c}}$ in Supplementary Table~\ref{Tabla1}). This figure shows (panel \textbf{a}) as a dashed blue line the symmetrical Gaussian peak used in our analyses (\textit{i.e.}, in Supplementary Fig.~\ref{Figura-EJ_rho-LSCO_0p12}b), while the solid red line is obtained by adding to this symmetrical distribution a small tail towards higher temperatures. In both distributions the main part of the peak is centered at the same $T_{\mathrm{cond}}$ value as in Supplementary Fig.~\ref{Figura-EJ_rho-LSCO_0p12}b, and also has the same FWHM $\Delta T_{\mathrm{c}}$=1.4K. The tail is built by adding for the upper temperatures a second Gaussian peak centered again at the same temperature as the main peak, but with much smaller amplitude (1/10 of the main peak) and much larger FWHM ($\Delta T\mathrm{_{c}^{\mathrm{tail}}}$=14K). Adding this upper tail produces a distribution that is almost indistinguishable from the symmetrical one, unless we zoom-in the upper temperature region as done in the inset of the Supplementary Fig.~\ref{Figura-Asymmetric_Gauss-LSCO_0p12}a. The area between both distributions within this region (grey shaded area, only easily noticed in the inset) accounts for merely about 4.5\% of the superconductor and is certainly well below the threshold of sensitivity in the SQUID measurements of the $T_{\mathrm{c}}$ distributions.

The Supplementary Fig.~\ref{Figura-Asymmetric_Gauss-LSCO_0p12}b compares the $\rho(T)$ measurements with a fit using the same effective-medium BKT/GGL approach as in our main text, but now employing $\varepsilon^{\mathrm{c}}=0.55$ and the critical temperature distribution with the upper asymmetric tail. The agreement with the data is excellent, at the same level as in Supplementary Fig.~\ref{Figura-EJ_rho-LSCO_0p12}b (that uses $\varepsilon^{\mathrm{c}}=0.9$ and the symmetric distribution). Importantly,
both the symmetric and asymmetric-distribution fits produce the same values for the average $T_{\mathrm{cond}}$. They also use the same average $T_{\mathrm{phase}}$ and also the same $b_{0}$, $\varepsilon_{\mathrm{LG}}$ and background. Therefore, the uncertainty in the precise value of $\varepsilon^{\mathrm{c}}$ does not affect the main conclusions of the present work, that concern the location of $T_{\mathrm{cond}}$ and $T_{\mathrm{phase}}$. This seems to be coherent with the fact that due to its physical meaning $\varepsilon^{\mathrm{c}}$ may be expected to influence the fit only well above both temperatures. 

We have checked that the same conclusions apply also for the rest of the samples displaying $\varepsilon^{\mathrm{c}}>0.55$ in Supplementary Table~\ref{Tabla1} (in particular, the tail for $\mathrm{LS{}_{0.11}CO}$ comprises 4\% of the sample and has $\Delta T\mathrm{_{c}^{\mathrm{tail}}}$=13K, and the one for $\mathrm{LS{}_{0.13}CO}$ comprises 2\% of the sample and has $\Delta T\mathrm{_{c}^{\mathrm{tail}}}$=7K).

%%%%%%%%%%%%%%%%%%%%%%%%%%%%%%%%%%%%%%%%%%%%%%%%%%%%%%
\begin{figure}[t]
\begin{centering}
\includegraphics[width=0.75\columnwidth]{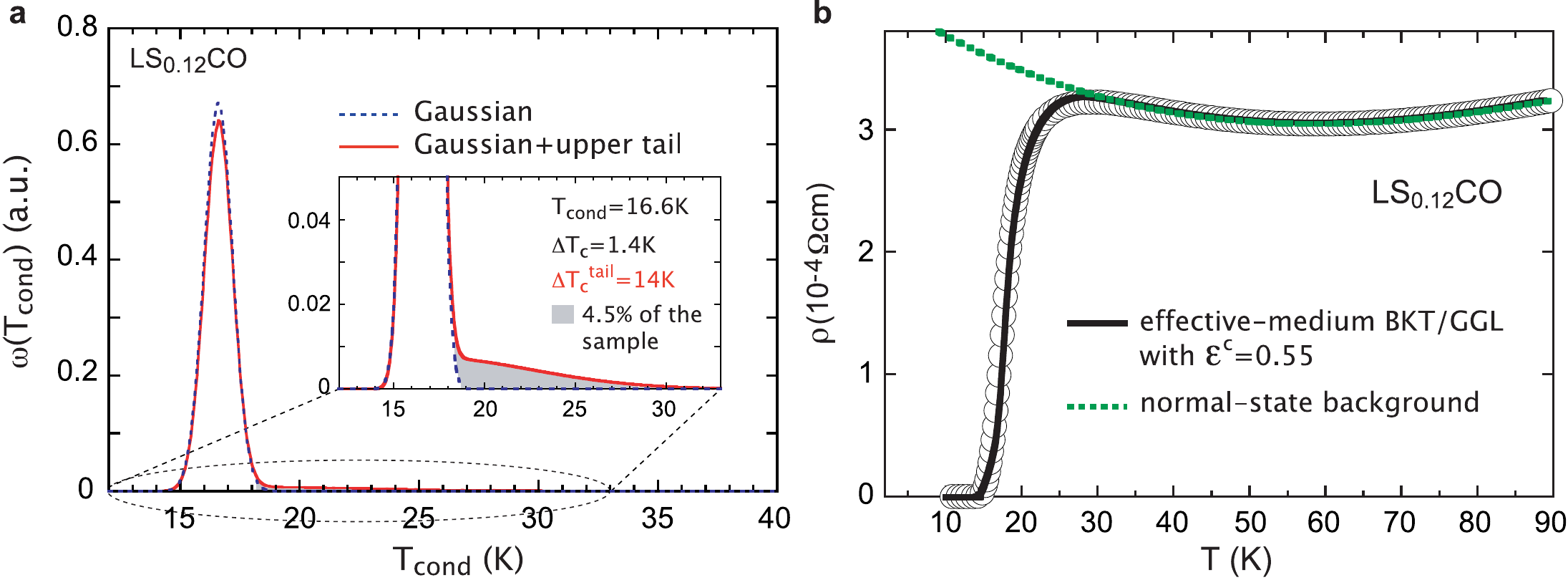}
\par\end{centering}
\caption{\label{Figura-Asymmetric_Gauss-LSCO_0p12} \small\sf Panel \textbf{a} shows the construction of the asymmetric critical temperature distribution mentioned in \ref{sec:AdditionalConsiderations}B, for sample $\mathrm{LS{}_{0.12}CO}$. The blue dashed line corresponds to a symmetric Gaussian distribution with the same average $T_{\mathrm{cond}}$ and $\Delta T_{\mathrm{c}}$ as in Supplementary Fig.~\ref{Figura-EJ_rho-LSCO_0p12}b. The red solid line corresponds to the same distribution plus an upper Gaussian tail centered at the same $T_{\mathrm{\mathrm{c}ond}}$ but with a larger FWHM, $\Delta T\mathrm{_{c}^{\mathrm{tail}}}$=14K. Panel \textbf{b} shows the results of repeating for this sample the same effective-medium BKT/GGL analyses as in Supplementary Fig.~\ref{Figura-EJ_rho-LSCO_0p12}b but now using the BCS clean-limit value of $\varepsilon^{\mathrm{c}}$ (\textit{i.e.}, $\varepsilon^{\mathrm{c}}$=0.55) and the asymmetric distribution of panel \textbf{a}.}
\end{figure}
%%%%%%%%%%%%%%%%%%%%%%%%%%%%%%%%%%%%%%%%%%%%%%%%%%%%%%

%%%%%%%%%%%%%%%%%%%%%%%%%%%%%%%%%%%%%%%%%%%%%%%%%%%%%%

\mbox{}

\textbf{C. \ Comparison with previous analyses of $\Delta\sigma$ in $\mathrm{L}\mathrm{a}_{2-x}\mathrm{S}\mathrm{r}_{x}\mathrm{Cu}\mathrm{O}_{4}$ as a function of doping}

The novel experimental elements in our present study allow improvements with respect to the previous investigations (see, \textit{e.g.}, Refs.\citeabajo{SuzukiHikita1991} to\citeabajo{BollingerDubuisetal2011}) of the paraconductivity $\Delta\sigma$ in cuprates with different dopings. In this regard, the three main new ingredients  are: The determination of $T_{\mathrm{phase}}$ by means of the $\alpha=3$ condition, the consideration of the critical temperature inhomogeneities (complemented with the values for $\Delta T_{\mathrm{\mathrm{c}}}$ resulting from independent high-precision magnetometry measurements), and the study of temperatures both above and below $T_{\mathrm{cond}}$. It may be useful to briefly comment here on the resulting advancements over some of the previous studies of $\Delta\sigma$ in $\mathrm{L}\mathrm{S}_{x}\mathrm{CO}$ as a function of doping.

We start with the previous work of our own group\citeabajo{CurrasFerroetal2003}, in which $\Delta\sigma$ was analyzed in $\mathrm{L}\mathrm{S}_{x}\mathrm{CO}$ films with $0.10\lesssim x\lesssim0.25$. In that study, the uncertainty (due to inhomogeneity effects) in the value of $T_{\mathrm{cond}}$ did not allow us to discard scenarios in which the superconducting fluctuations were outside the 2D limit.\renewcommand{\thefootnote}{\fnsymbol{footnote}}\footnotemark[2]\footnotetext[2]{\small\sf We have checked that analyzing our present $\Delta\sigma$ data using an scenario outside the 2D limit for the GGL regime of  fluctuations also does not lead to $T_{\rm cond}$ values larger than $T_{90\%}$, so that the main result from our present $\Delta\sigma$ analyses would remain valid even in that scenario. However, the consistency with the  data would be significantly worse than in the 2D limit.} Also, these analyses were limited to the temperatures above $T_{\mathrm{cond}}$ (in fact, above $\mathrm{\varepsilon}_{\mathrm{LG}}$), not entering into the BKT-like region of the fluctuations nor leading to any proposal for $T_{\mathrm{phase}}$ (for comparison, the region analyzed in that work would correspond to only $t\gtrsim0.2$ in the Supplementary Fig.~\ref{Figura-paraconductividad}). Let us also mention that our present work reduces the uncertainty in the background subtraction by using the \textit{$T$}-dependence of $\mathrm{d}\rho/\mathrm{d}T$ instead of the $\rho(T)$ one; even so, the central-value backgrounds used in\citeabajo{CurrasFerroetal2003} lead to $\Delta\sigma$ results that in the region $0.02\lesssim\varepsilon\lesssim0.1$ essentially agree with those obtained in the present work.

More recently, Rourke and coworkers studied in\citeabajo{Rourke} the paraconductivity in $\mathrm{L}\mathrm{S}_{x}\mathrm{CO}$ bulk crystals of the overdoped compositions $0.21\lesssim x\lesssim0.26$ (see also the Comment\citeabajo{commentRourke}). No \textit{$V-I$} curves or $T_{\mathrm{c}}$ distributions were measured or taken into
account in that work. These authors claim $T_{\mathrm{cond}}$ values well larger than $\mathrm{\mathit{T}_{phase}}$, by presenting analyses in which they propose the identification of $T_{\mathrm{cond}}$ with the temperature of the upturn of $\mathrm{d}\rho/\mathrm{d}T$ with respect to \textit{$T$}, \textit{i.e.}, in our notation, $T_{\mathrm{cond}}=T_{\mathrm{upturn}}$. Also, the background used to obtain $\Delta\sigma$ is extracted from the temperatures immediately above $T_{\mathrm{upturn}}$. Note however that the latter two ansatzs are not really compatible with each other: The proposal of superconducting fluctuations strong enough to renormalize the critical temperature about 20 to 60K (the typical range of values for $T_{\mathrm{\mathrm{con}d}}-T_{\mathrm{phase}}$ proposed in\citeabajo{Rourke}) is in disagreement with the proposal that those fluctuations would be negligible right above $T_{\mathrm{cond}}$. Obviously this is also in contradiction with the calculations: Direct application of equation~(\ref{eq:GGL}) for $\varepsilon=0.1$, \textit{i.e.}, $T\simeq1.1T_{\mathrm{cond}}$ (corresponding for instance to about 6K above $T_{\mathrm{upturn}}$ for the LS$_{0.21}$CO sample in\citeabajo{Rourke}) leads to GGL fluctuations already of the order of $\Delta\sigma\sim10^{5}$~$(\Omega\mathrm{m)}^{-1}$,\renewcommand{\thefootnote}{\fnsymbol{footnote}}\footnotemark[3]\footnotetext[3]{\small\sf At $\varepsilon=0.1$  cutoff effects may still be neglected for any $\varepsilon^{\rm c}\stackrel{>}{_\sim}0.3$, and so this estimate essentially holds for any practical value of $\varepsilon^{\rm c}$.} which would certainly produce well visible deviations (of at least 10\%) from the resistivity background, contrarily to the proposals in\citeabajo{Rourke}. Moreover, the ensuing analyses of $\Delta\sigma$ below $T_{\mathrm{upturn}}$ lead the authors of\citeabajo{Rourke} to propose a value of $T_{\mathrm{phase}}$ similar to the inflection point of the $\rho(T)$ macroscopic transition. As we have shown by means of the more unambiguous $\alpha=3$ condition in the \textit{$E-J$} curves, $T_{\mathrm{phase}}$ must actually be lower. Regardless of these shortcomings, note also that Rourke and coworkers fit the expression $\Delta\sigma\propto\sinh^2\left[b_{0}\Delta_{\mathrm{BKT}}/(T-T_{\mathrm{phase}})\right]^{1/2}$ to their proposed paraconductivity below $T_{\mathrm{upturn}}$. Although they claim this fit to be indicative of BKT behaviour, that expression is actually a formula first proposed by Halperin and Nelson\cite{HN} to interpolate   the paraconductivity from the BKT exponential behaviour to a mean-field-like law $\stackrel{\propto}{_\sim}t^{-1}$. The expression includes then both a BKT-like and a GGL-like regime. Given that for any temperature between $T_{\mathrm{phase}}$ and $T_{\mathrm{cond}}$ the full-critical renormalization must be already taking place, the $T_{\mathrm{phase}}<T<T_{\mathrm{cond}}$ range must be into the exponential part of the interpolation formula.\cite{HN} In contrast, in\citeabajo{Rourke} this is not the case: For instance, using the fitting parameter values proposed for  LS$_{0.21}$CO in\citeabajo{Rourke}, which includes $b_{0}=0.01$, the mean-field-like behaviour actually holds for all the analyzed data, down to $T_{\mathrm{phase}}+$0.5K (much below  than the $T_{\mathrm{cond}}$ value claimed in\citeabajo{Rourke} for that sample, $T_{\mathrm{cond}}\simeq T_{\mathrm{phase}}+$30K).

Finally, some very recent papers\cite{ShiLogvenovetal2013,BollingerDubuisetal2011}  study the electric transport in the superconducting-to-insulating transition characteristic of the low-doping boundary of the superconducting dome in $\mathrm{L}\mathrm{S}_{x}\mathrm{CO}$ films (grown over $\mathrm{LaSrAlO_{4}}$ substrates, known to increase the critical temperatures\cite{SatoNaito1997,BozovicLogvenovetal2002}). The influence of the quantum fluctuation effects near $T=0$K, and of the likely inhomogeneous antiferromagnetic cluster spin glass ordering, makes debatable the adequateness of directly comparing the results in\citeabajo{ShiLogvenovetal2013,BollingerDubuisetal2011} to those in the present work. Even so, in\citeabajo{ShiLogvenovetal2013} it is claimed that for $0.03\lesssim x\lesssim0.08$ the superconducting fluctuations have  onset temperatures lower by at least about 10 to 20K than claimed for the strong phase fluctuations scenario.\cite{Franz,Emery,AndersonNat,Lascialfari2002,OngNeruno,OngMagPRB} A similar result was in fact obtained in\citeabajo{BilbroValdesetal2011} for similar samples with dopings $0.09\lesssim x\lesssim0.25$,  using conductivity measurements at THz frequency (that could in principle change the nature of the fluctuations by competing with their relaxation time\cite{RamalloCarballeiraetal1999}). These relatively low onset temperatures seem to be compatible with our findings for $T_{\mathrm{cond}}-T_{\mathrm{phase}}$. In contrast, a larger regime of strong phase fluctuations was suggested in\citeabajo{BollingerDubuisetal2011} using scaling arguments of $\rho$ at low $T$ measured in $\mathrm{L}\mathrm{S}_{x}\mathrm{CO}$/$\mathrm{LaSrAlO_{4}}$ films with variable doping induced by electric fields. It is unclear whether this is related to the fact that samples in\citeabajo{BollingerDubuisetal2011} are only 1 to 5 monolayer thick, what could enhance their phase fluctuations.

%%%%%%%%%%%%%%%%%%%%%%%%%%%%%%%%%%%%%%%%%%%%%%%%%%%%%%
%%%%%%%%%%%%%%%%%%%%%%%%%%%%%%%%%%%%%%%%%%%%%%%%%%%%%%
%%%%%%%%%%%%%%%%%%%%%%%%%%%%%%%%%%%%%%%%%%%%%%%%%%%%%%
%%%%%%%%%%%%%%%%%%%%%%%%%%%%%%%%%%%%%%%%%%%%%%%%%%%%%%
%%%%%%%%%%%%%%%%%%%%%%%%%%%%%%%%%%%%%%%%%%%%%%%%%%%%%%
%%%%%%%%%%%%%%%%%%%%%%%%%%%%%%%%%%%%%%%%%%%%%%%%%%%%%%

\renewcommand\refname{Supplementary References}\clearpage{}\newpage{}

\setlength\bibsep{0.4em}
\begin{thebibliography}{S99}

\bibitem[S1]{RamalloeLG}Ramallo, M.V. \& Vidal, F. On the width of the full-critical region for thermal fluctuations around the superconducting transition in layered superconductors. {\it Europhys. Lett.} {\bf 39,} 177-182 (1997).

\bibitem[S2]{MosqueiraPtos}Mosqueira, J., Cabo, L. \& Vidal, F. Structural and $T_{\rm c}$ inhomogeneities inherent to doping in ${\rm La}_{2-x}{\rm Sr}_x{\rm CuO}_4$ superconductors and their effects on the precursor diamagnetism. {\it Phys. Rev. B} {\bf 80,} 214527 (2009).

\bibitem[S3]{Radaelli}Radaelli, P.G. {\it et al.} Structural and superconducting properties of ${\rm La}_{2-x}{\rm Sr}_x{\rm CuO}_4$ as a function of Sr content. {\it Phys. Rev. B} {\bf 49,} 4163-4175 (1994).

\bibitem[S4]{Kofu}Kofu, M., Kimura, H. \& Hirota, K. Zn and Ni doping effects on the low-energy spin excitations in ${\rm La}_{1.85}{\rm Sr}_{0.15}{\rm CuO}_4$.{\it Phys. Rev. B} {\bf 72,} 064502 (2005).

\bibitem[S5]{Takagi}Takagi, H. {\it et al.} Superconductor-to-nonsuperconductor transition in $({\rm La}_{1-x}{\rm Sr}_x)_2{\rm CuO}_4$ as investigated by transport and magnetic measurements. {\it Phys. Rev. B} {\bf 40,} 2254-2261 (1989).

\bibitem[S6]{Torrance}Torrance, J.B. {\it et al.} Anomalous Disappearance of High-$T_{\rm c}$ Superconductivity at High Hole Concentration in Metallic. {\it Phys. Rev. Lett.} {\bf 61,} 1127-1130 (1988).

\bibitem[S7]{Zhou}Zhou, F. {\it et al.} Under-doped  ${\rm La}_{2-x}{\rm Sr}_x{\rm Cu}{\rm O}_4$ with $x$= 0.063\textendash{}0.125: TSFZ growth of high-quality crystals and anomalous doping dependences of superconducting properties. {\it Supercond. Sci. Technol.} {\bf 16,} L7-L9 (2003).

\bibitem[S8]{Marin}Marin, C., Charvolin, T., Braithwaite, D. \& Calemczuk, R. Properties of a large ${\rm La}_{1.92}{\rm Sr}_{0.08}{\rm CuO}_{4+\delta}$ single crystal grown by the travelling-solvent floating-zone method. {\it Physica C} {\bf 320,} 197-205 (1999).

\bibitem[S9]{MazaEM}Maza, J. \& Vidal, F. Critical-temperature inhomogeneities and resistivity rounding in copper oxide superconductors. {\it Phys. Rev. B} {\bf 43,} 10560-10567 (1991).

\bibitem[S10]{CotonHN2}Cot\'on, N. {\it et al.} Finite-Element and Effective-Medium Calculations of the Electrical Behaviour Near the Vortex-Antivortex Binding Transition of Planar Superconductors with Critical Temperature Inhomogeneities. {\it J. Supercond. Nov. Magn.} {\bf 26,} 3065-3068 (2013).

\bibitem[S11]{SuzukiHikita1991}Suzuki, M. \& Hikita, M. Resistive transition, magnetoresistance, and anisotropy in ${\rm La}_{2-x}{\rm Sr}_x{\rm Cu}{\rm O}_4$ single-crystal thin films. {\it Phys. Rev. B} {\bf 44}, 249-261 (1991).

\bibitem[S12]{SuzukiHikita1993}Suzuki, M. \& Hikita, M. Fluctuation conductivity and normal resistivity in ${\rm La}_{2-x}{\rm Sr}_x{\rm Cu}{\rm O}_4$ as a function of doping. {\it Phys. Rev. B} {\bf 47}, 2913-2915 (1993).

\bibitem[S13]{KimuraMiyasakaetal1996}Kimura, T. {\it et al.} In-plane and out-of-plane magnetoresistance in ${\rm La}_{2-x}{\rm Sr}_x{\rm Cu}{\rm O}_4$ single crystals. {\it Phys. Rev. B} {\bf 53}, 8733-8742 (1996).

\bibitem[S14]{CurrasFerroetal2003}Curr\'as, S.R. {\it et al.} In-plane paraconductivity in ${\rm La}_{2-x}{\rm Sr}_x{\rm Cu}{\rm O}_4$ thin film superconductors at high reduced temperatures: Independence of the normal-state pseudogap. {\it Phys. Rev. B} {\bf 68}, 094501 (2003).

\bibitem[S15]{LeridonVanackenetal2007}Leridon, B., Vanacken, J., Wambecq, T. \& Moshchalkov, V.V. Paraconductivity of underdoped ${\rm La}_{2-x}{\rm Sr}_x{\rm Cu}{\rm O}_4$ thin-fi{}lm superconductors using high magnetic fields. {\it Phys. Rev. B} {\bf 76}, 012503 (2007).

\bibitem[S16]{Rourke}Rourke, P.M. {\it et al.} Phase-fluctuating superconductivity in overdoped ${\rm La}_{2-x}{\rm Sr}_x{\rm Cu}{\rm O}_4$. {\it Nature Physics} {\bf 7},
455-458 (2011).

\bibitem[S17]{commentRourke}Mosqueira, J., Ramallo, M.V. \& Vidal, F. Comment on ``Phase fluctuation superconductivity in overdoped ${\rm La}_{2-x}{\rm Sr}_x{\rm Cu}{\rm O}_4$''. http://arxiv.org/abs/1112.6104v1 (2011).

\bibitem[S18]{ShiLogvenovetal2013}Shi, X. {\it et al.} Emergence of superconductivity from the dynamically heterogeneous insulating state in ${\rm La}_{2-x}{\rm Sr}_x{\rm Cu}{\rm O}_4$. {\it Nature Materials} {\bf 12}, 47-51 (2013).

\bibitem[S19]{BollingerDubuisetal2011}Bollinger, A.T. {\it et al.} Superconductor\textendash{}insulator transition in ${\rm La}_{2-x}{\rm Sr}_x{\rm Cu}{\rm O}_4$ at the pair quantum resistance. {\it Nature} {\bf 472}, 458-460 (2011).

\bibitem[S20]{SatoNaito1997}Sato, H. \& Naito, M. Increase in the superconducting transition temperature by anisotropic strain effect in $(001){\rm La}_{1.85}{\rm Sr}_{0.15}{\rm Cu}{\rm O}_4$ thin films on ${\rm LaSrAlO}_4$ substrates. {\it Physica C} {\bf 274}, 221-226 (1997).

\bibitem[S21]{BozovicLogvenovetal2002}Bo\v{z}ovi\'c, I., Logvenov, G., Belca, I., Narimbetov, B. \& Sveklo, I. Epitaxial Strain and Superconductivity in ${\rm La}_{2-x}{\rm Sr}_x{\rm Cu}{\rm O}_4$ Thin Films. {\it Phys. Rev. Lett.} {\bf 89}, 107001 (2002).

\bibitem[S22]{BilbroValdesetal2011}Bilbro, L.S. {\it et al.} Temporal correlations of superconductivity above the transition temperature in ${\rm La}_{2-x}{\rm Sr}_x{\rm Cu}{\rm O}_4$ probed by terahertz spectroscopy. {\it Nature Physics} {\bf 7}, 298-302 (2011).

\bibitem[S23]{RamalloCarballeiraetal1999}Ramallo, M.V. {\it et al.} Relaxation time of the Cooper pairs near $T_{\mathrm{c}}$ in cuprate superconductors. {\it Europhys. Lett.} {\bf 48}, 79-85 (1999).

\endgroup  %%<-- neccesary hack to allow two thebibliography sections in one latex document

\end{document}